\documentclass[prb,longbibliography,twocolumn,eqsecnum,nofootinbib,reprint]{revtex4}
\usepackage[cp850]{inputenc}
\usepackage{natbib}
\usepackage{amssymb,amsmath}
\usepackage{epsfig}
\usepackage{bm}
\usepackage{color}

\usepackage{graphicx}

\usepackage{color}
\newcommand{\vicente}[1]{{ #1}}

\newcommand\beq{\begin{equation}}
\newcommand\eeq{\end{equation}}
\newcommand\beqa{\begin{eqnarray}}
\newcommand\eeqa{\end{eqnarray}}
\newcommand{\dd}{\text{d}}

\newcommand{\al}{\alpha}

\begin{document}

\title{Non-Newtonian rheology in inertial suspensions of inelastic rough hard spheres under simple shear flow}

\author{Rub\'en G\'omez Gonz\'alez\footnote[1]{Electronic address: ruben@unex.es}}
\affiliation{Departamento de F\'{\i}sica,
Universidad de Extremadura, E-06006 Badajoz, Spain}
\author{Vicente Garz\'{o}\footnote[2]{Electronic address: vicenteg@unex.es;
URL: http://www.unex.es/eweb/fisteor/vicente/}}
\affiliation{Departamento de F\'{\i}sica and Instituto de Computaci\'on Cient\'{\i}fica Avanzada (ICCAEx), Universidad de Extremadura, E-06006 Badajoz, Spain}

\begin{abstract}
Non-Newtonian transport properties of an inertial suspension of inelastic rough hard spheres under simple shear flow are determined from the Boltzmann kinetic equation. The influence of the interstitial gas on rough hard spheres is modeled via a Fokker--Planck generalized equation for rotating spheres accounting for the coupling of both the translational and rotational degrees of freedom of grains with the background viscous gas. The generalized Fokker--Planck term is the sum of two ordinary Fokker--Planck differential operators in linear $\mathbf{v}$ and angular $\boldsymbol{\omega}$ velocity space. As usual, each Fokker--Planck operator is constituted by a drag force term (proportional to $\mathbf{v}$ and/or $\boldsymbol{\omega}$) plus a stochastic Langevin term defined in terms of the background temperature $T_\text{ex}$. The Boltzmann equation is solved by two different but complementary approaches: (i) by means of Grad's moment method, and (ii) by using a Bhatnagar--Gross--Krook (BGK)-type kinetic model adapted to inelastic rough hard spheres. As occurs in the case of \emph{smooth} inelastic hard spheres, our results show that both the temperature and the non-Newtonian viscosity increase drastically with increasing the shear rate (discontinuous shear thickening effect) while the fourth-degree velocity moments also exhibit an $S$-shape. In particular, while high levels of roughness may slightly attenuate the jump of the viscosity in comparison to the smooth case, the opposite happens for the rotational temperature. As an application of these results, a  linear stability analysis of the steady simple shear flow solution is also carried out showing that there are regions of the parameter space where the steady solution becomes linearly unstable. The results derived here extend to rough spheres previous works devoted to purely smooth spheres [H. Hayakawa and S. Takada, Prog. Theor. Exp. Phys. \textbf{083J01} (2019); R. G\'omez Gonz\'alez and V. Garz\'o, J. Stat. Mech. \textbf{013206} (2019)].
\end{abstract}

\draft
\date{\today}
\maketitle

\section{Introduction}
\label{sec1}

Needless to say, shear thickening (a rheological process in which the viscosity increases with the shear rate) in non-Newtonian gas-solid flows is likely one of the most challenging and open problems in suspensions of particles in gases or liquids. Apart from its practical interest (it has been broadly found in nature\cite{C97a} and industry \cite{ChOLS11,PBChDHILC18}), its understanding from a more fundamental point of view has attracted the attention of many researchers in the last few years. \cite{B89,LDHH05,BJ09,MW11,CPNC11,OH11,H13,SMMD13,BJ14,HJGGLS17,KFZFS18,MOH18,ChMCF18,HRZSI18,SPChDM19,JB19,RBU20,SNSPJ20} Shear thickening can occur as a smooth increase of the viscosity with increasing the shear rate; this effect is usually referred to as continuous shear thickening (CST). On the other hand, it can be also observed as a drastic increase of the viscosity at a specific shear rate; this dramatic version of CST is known as discontinuous shear thickening (DST). These two different phenomena can be observed for instance in a suspension of cornstarch on water at different cornstarch concentrations.

On the other hand, although the shear-induced solid-like behavior produced in DST has generated a significant interest, most of the studies have been focused in densely packed suspensions where extensive simulations have been carried out to disclose the origin of this unexpected phenomenon. As has been widely discussed in the review of Brown and Jaeger,\cite{BJ14} the above studies propose three main mechanisms based on particle reorganization to explain the shear thickening phenomena: hydroclustering, order-disorder transition, and/or dilatancy. However, DST has been shown to appear also at relatively low-density regimes\cite{TK95,SMTK96,ChVG15,SA17,SA20,HTG17,HTG20,HT19,GGG19} where specific structural characteristics that influence the stress transmission are not apparently substantial enough to explain such a sharp transition. Thus, in order to unveil in a clean way the microscopic mechanisms involved in DST, it would be also convenient to consider relatively low-density systems where kinetic theory can provide a quantitative theoretical description. In the context of kinetic theory, some previous works \cite{TK95,SMTK96,ChVG15,SA17,SA20} have shown the existence of a DST-like process for the temperature between a \emph{quenched} state (a low-temperature state) and an \emph{ignited} state (a high-temperature state) in homogeneously sheared gas-solid suspensions.

However, all the above works \cite{TK95,SMTK96,ChVG15,SA17,SA20} consider a suspension model where the effects of thermal fluctuations on the dynamics of grains were neglected. A more accurate suspension model where the effect of the interstitial gas on solid particles is accounted for via a viscous drag force plus a stochastic Langevin term \cite{GTSH12} has been recently considered \cite{HTG17,HTG20,HT19} for obtaining the shear-rate dependence of the kinetic temperature and the stress tensor. The theoretical results \cite{HTG17,HT19,HTG20,GGG19} have been compared against event-driven Langevin simulation for hard spheres (EDLSHS), \cite{S12} showing a very good agreement specially for low-density systems. Both approaches (kinetic theory and simulations) conclude that there is a transition from DST (found for very dilute systems) to CST as the volume fraction of the granular gas increases.

An important limitation of the above theoretical works \cite{TK95,SMTK96,ChVG15,SA17,SA20,HTG17,HTG20,HT19,GGG19} is that the solid particles were modeled as \emph{smooth} inelastic hard spheres. This means that the effects of tangential friction and rotation induced by each binary collision on rheology were ignored in the above attempts. The purpose of the present paper is to extend the previous theoretical efforts of smooth spheres to \emph{rough} spheres in order to assess the impact of roughness on the rheological properties of the suspension. Thus, we want to uncover the whole range values of the normal $\al$ and tangential $\beta$ restitution coefficients and derive explicit expressions for the rotational $T_r$ and translational $T_t$ temperatures as well as for the relevant elements of the pressure tensor $P_{k\ell}$. Given the mathematical difficulties involved in the general problem,  as in Refs.\ \onlinecite{HT19,GGG19}, we consider here very dilute systems for which the Boltzmann kinetic equation offers a reliable description. To the best of our knowledge, only \vicente{three} previous papers \cite{LDHH05,HJGGLS17,HRZSI18} have addressed the role of roughness in the rheological phenomena. However, given that these works \cite{LDHH05,HJGGLS17,HRZSI18} consider concentrated colloidal suspensions at the jamming transition, no analytical results were derived since they combine experimental and computer simulation results of spherical colloids. In this sense, the present contribution complements these previous attempts \cite{LDHH05,HJGGLS17,HRZSI18} since our results allows us to unveil the combined effect of both $\al$ and $\beta$ on the shear-rate dependence of the pressure tensor.

As said before, our goal here is to determine the rheological properties of an inertial suspension of inelastic rough hard spheres under simple shear flow. This state is macroscopically characterized by a constant density $n$, a uniform temperature $T$, and an homogeneous shear field $U_x=ay$, where $a$ is the constant shear rate. As usual, we are interested here in steady state conditions. In addition, as in previous works, \cite{HT19,GGG19} the influence of the viscous gas on solid particles is modeled by means of an operator representing the gas-solid interaction force. In the limit case of purely smooth spheres ($\al=1$ and $\beta=-1$), only translational degrees of freedom play a role in the dynamics of grains. In this special case, the fluid-force is composed by a viscous drag force proportional to the (instantaneous) velocity of particles $\mathbf{v}$ (the coefficient of proportionality is the translational drift coefficient $\gamma_t$) plus a Langevin-like term defined in terms of the background temperature $T_\text{ex}$. On the other hand, beyond the smooth case, one has to take into account the coupling between the rotational degrees of freedom of grains and the interstitial gas. Following a model introduced years ago by Hess \cite{H68} for Brownian motion of rotating particles, we assume that the structure of the rotational part of the fluid-force is similar to that of the translational part: a drag force term proportional to the angular velocity $\boldsymbol{\omega}$ (the coefficient of proportionality is the rotational drift coefficient $\gamma_r$) plus a stochastic Langevin-like term defined in terms of $T_\text{ex}$. The coefficients $\gamma_t$ and $\gamma_r$ are both proportional to the shear viscosity of the interstitial gas and hence, both coefficients are proportional to $\sqrt{T_\text{ex}}$. This suspension model has been more recently considered to study a segregation problem of microswimmer mixtures. \cite{JLHBL19}

The suspension model for inelastic rough hard spheres is solved by following two different but complementary theoretical tools. First, Grad's moment method\cite{G49} is considered to \emph{approximately} get the explicit forms of both the (reduced) translational $T_t/T_\text{ex}$ and rotational $T_r/T_\text{ex}$ temperatures and the (reduced) elements $P_{k\ell}/(nT_\text{ex})$ of the pressure tensor in terms of the restitution coefficients $\al$ and $\beta$ and the (reduced) shear rate $a^*\equiv a/\gamma_t$. Then, as a second alternative and to overcome the mathematical difficulties of the Boltzmann collision operator, a Bhatnagar-Gross-Krook (BGK) model kinetic equation recently proposed for inelastic rough hard spheres \cite{S11} is considered. This kinetic model retains the essential physical properties of the Boltzmann equation and allow one to obtain all the velocity moments of the velocity distribution function. In particular, the results derived for the pressure tensor from the kinetic model coincide with those derived from the Boltzmann equation when one conveniently chooses a free parameter of the model. Apart from the second-degree velocity moments, the shear-rate dependence of the fourth-degree moments is also widely analyzed.

The plan of the paper is as follows. Section \ref{sec2} is devoted to the definition of the suspension model for inelastic rough hard spheres in the low-density limit. Starting from the Boltzmann kinetic equation, the exact balance equations for the densities of mass, momentum, and energy are derived with expressions for the momentum and heat fluxes. These expressions are defined in terms of the velocity distribution function. Section \ref{sec3} deals with the simple shear flow state where the time evolution of the elements of the pressure tensor $P_{k\ell}$ is exactly obtained. The above set of equations for $P_{k\ell}$ is solved  by estimating the collisional moment associated with the transfer of momentum by means of Grad's moment method. This permits to achieve explicit forms for $T_r$, $T_t$, and $P_{k\ell}$ under steady state conditions. The results obtained from the BGK-like model are exposed in Sec.\ \ref{sec4}. Before considering the results for inertial suspensions, Sec.\ \ref{sec5} analyzes the results in the so-called \emph{dry} granular gases, namely, when the influence of the interstitial gas is neglected (i.e., when $\gamma_t=\gamma_r=0$). Although these results are interesting by themselves, they offer the opportunity to compare the present theory with the results derived many years ago by Lun \cite{L91} for nearly elastic collisions ($\al \lesssim 1$) and nearly perfectly rough particles ($\beta\lesssim 1$). The results for the rheological properties and the fourth-degree velocity moments of inertial suspensions are illustrated in Sec.\ \ref{sec6} for several values of the coefficients $\al$ and $\beta$. It is clearly shown that the roughness do not substantially change the conclusions found in the smooth limit case since DST is also present for inelastic rough spheres. In addition, the BGK results also show that the fourth-degree moments increase dramatically with the shear rate in a certain region of values of the shear rate. A linear stability analysis of the steady simple shear flow solution is carried out in Sec.\ \ref{sec7}. As expected from the previous analysis performed for smooth spheres, \cite{HT19} the homogeneous steady sheared solution can be linearly \emph{unstable} in certain regions of the parameter space. The paper is closed in Sec.\ \ref{sec8} with a brief discussion on the results reported here.

\section{Boltzmann kinetic equation for gas-solid flows of inelastic rough hard spheres}
\label{sec2}

\subsection{Boltzmann equation for inertial suspensions}

We consider a set of solid particles of diameter $\sigma$, mass $m$, and moment of inertia $I$ immersed in a molecular gas of viscosity $\eta_g$. The solid particles are modeled as inelastic rough hard spheres. We assume that the collisions among particles are inelastic and are characterized by constant coefficients of normal restitution ($\alpha$) and tangential restitution ($\beta$). While the coefficient $\al$ ranges from 0 (perfectly inelastic collisions) to 1 (perfectly elastic collisions), the coefficient $\beta$ ranges from $-1$ (perfectly smooth spheres) to $1$ (perfectly rough spheres). Kinetic energy is in general dissipated by collisions, except in the cases $\al=1$ and $\beta=\pm 1$. An interesting feature of this model is that inelasticity affects both translational and rotational degrees of freedom of the spheres.

In the low-density regime ($n\sigma^3\ll 1$, where $n$ is the number density), all the relevant information on the state of the suspension is given through the one-particle velocity distribution function $f(\mathbf{r}, \mathbf{v}, \boldsymbol{\omega}; t)$, where $\mathbf{v}$ and $\boldsymbol{\omega}$ are the (instantaneous) linear (translational) and angular velocities, respectively. Neglecting the effects of the gravity field, the velocity distribution $f$ obeys the Boltzmann kinetic equation \cite{JR85a,GS95,Z06,BP04,G19}
\beq
\label{2.1}
\frac{\partial f}{\partial t}+\mathbf{v}\cdot \nabla f+\mathcal{F} f=J\big[\mathbf{v}, \boldsymbol{\omega}|f(t),f(t)\big],
\eeq
where $\mathcal{F} f$ is an operator characterizing the influence of the interstitial gas on grains and $J[f,f]$ is the Boltzmann collision operator given by \cite{BP04,G19}
\beqa
\label{2.2}
& & J[\mathbf{v}_1, \boldsymbol{\omega}_1|f,f]=\sigma^{2}\int \dd{\bf v}_{2}\int \dd\boldsymbol{\omega}_2 \int \dd\widehat{\boldsymbol{\sigma}} \nonumber\\
& & \times \Theta (\widehat{{\boldsymbol {\sigma}}}\cdot {\bf g})(\widehat{\boldsymbol {\sigma }}\cdot {\bf g})\Big[\frac{1}{\alpha^2\beta^2}f(\mathbf{r},\mathbf{v}_1'', \boldsymbol{\omega}_1'';t)f(\mathbf{r},\mathbf{v}_2'', \boldsymbol{\omega}_2'';t)
\nonumber\\
& &
-f(\mathbf{r},\mathbf{v}_1, \boldsymbol{\omega}_1; t)f(\mathbf{r},\mathbf{v}_2, \boldsymbol{\omega}_2; t)\Big].
\eeqa
Here, $\Theta(x)$ is Heaviside's step function, $\boldsymbol{\widehat{\sigma}}$ is the unit collision vector joining the centers of the two colliding spheres and pointing from the sphere labeled by 1 to the sphere labeled by 2, and $\mathbf{g}=\mathbf{v}_1-\mathbf{v}_2$ is the relative translational velocity. In Eq.\ \eqref{2.2}, the double primes on the linear and angular velocities denote the initial velocities $\left\{\mathbf{v}_1'',\boldsymbol{\omega}_1'', \mathbf{v}_2'', \boldsymbol{\omega}_2''\right\}$ that lead to the final velocities  $\left\{\mathbf{v}_1,\boldsymbol{\omega}_1, \mathbf{v}_2, \boldsymbol{\omega}_2\right\}$ following a binary restituting collision. The restituting (or \emph{inverse}) collision rules are \cite{JR85a,L91,GS95,Z06,SKG10}
\beq
\label{2.3}
\mathbf{v}_1''=\mathbf{v}_1-\mathbf{Q}'', \quad \mathbf{v}_2''=\mathbf{v}_2+\mathbf{Q}'',
\eeq
\beq
\label{2.4}
\boldsymbol{\omega}_1''=\boldsymbol{\omega}_1-\frac{2}{\sigma\kappa}\boldsymbol{\widehat{\sigma}}\times \mathbf{Q}'',
\quad \boldsymbol{\omega}_2''= \boldsymbol{\omega}_2-\frac{2}{\sigma\kappa}\boldsymbol{\widehat{\sigma}}\times \mathbf{Q}'',
\eeq
where $\mathbf{Q}''$ reads
\beqa
\label{2.5}
\mathbf{Q}''&=&\frac{1+\al^{-1}}{2}\boldsymbol{\widehat{\sigma}}(\boldsymbol{\widehat{\sigma}}\cdot \mathbf{g})-\frac{\kappa}{1+\kappa}\frac{1+\beta^{-1}}{2}\Big[\boldsymbol{\widehat{\sigma}}(\boldsymbol{\widehat{\sigma}}\cdot \mathbf{g})-\mathbf{g}\nonumber\\
& & +\frac{\sigma}{2}\boldsymbol{\widehat{\sigma}}\times (\boldsymbol{\omega}_1+\boldsymbol{\omega}_2)\Big].
\eeqa
In Eqs.\ \eqref{2.4} and \eqref{2.5}, $\kappa=4I/m\sigma^2$ is a dimensionless parameter characterizing the mass distribution within a sphere. It runs from the extreme values $\kappa=0$ (namely, when the mass is concentrated on the center of the sphere) and $\kappa=\frac{2}{3}$ (namely, when the mass is concentrated on the surface of the sphere). In the case that the mass is uniformly distributed, then $\kappa=\frac{2}{5}$.

Similarly, the collisional rules for the \emph{direct} collision $(\mathbf{v}_1, \boldsymbol{\omega}_1,\mathbf{v}_2, \boldsymbol{\omega}_2)\to (\mathbf{v}_1', \boldsymbol{\omega}_1',\mathbf{v}_2', \boldsymbol{\omega}_2')$ are
\beq
\label{2.13}
\mathbf{v}_1'=\mathbf{v}_1-\mathbf{Q}, \quad \mathbf{v}_2'=\mathbf{v}_2+\mathbf{Q},
\eeq
\beq
\label{2.13.1}
\boldsymbol{\omega}_1'=\boldsymbol{\omega}_1-\frac{2}{\sigma\kappa}\boldsymbol{\widehat{\sigma}}\times \mathbf{Q}, \quad
\boldsymbol{\omega}_2'=\boldsymbol{\omega}_2-\frac{2}{\sigma\kappa}\boldsymbol{\widehat{\sigma}}\times \mathbf{Q},
\eeq
where $\mathbf{Q}$ is given by
\beqa
\label{2.14}
\mathbf{Q}&=&\frac{1+\al}{2}\boldsymbol{\widehat{\sigma}}(\boldsymbol{\widehat{\sigma}}\cdot \mathbf{g})-\frac{\kappa}{1+\kappa}\frac{1+\beta}{2}\Big[\boldsymbol{\widehat{\sigma}}(\boldsymbol{\widehat{\sigma}}\cdot \mathbf{g})-\mathbf{g}\nonumber\\
& & +\frac{\sigma}{2}\boldsymbol{\widehat{\sigma}}\times (\boldsymbol{\omega}_1+\boldsymbol{\omega}_2)\Big].
\eeqa
Equations \eqref{2.13} and \eqref{2.14} allows us to evaluate the variation of the total energy (translational plus rotational energy). After some algebra, one gets
\begin{widetext}
\beqa
\label{2.14.1}
\Delta E&=&\frac{m}{2}\left(v_1^{'2}+v_2^{'2}-v_1^2-v_2^2\right)+\frac{I}{2}\Big(\omega_1^{'2}+\omega_2^{'2}-\omega_1^{2}
-\omega_2^{2}\Big)\nonumber\\
&=&-m \frac{1-\beta^2}{4}\frac{\kappa}{1+\kappa}\Big[\boldsymbol{\widehat{\sigma}}\times \Big(\boldsymbol{\widehat{\sigma}}\times \mathbf{g}
+\sigma \frac{\boldsymbol{\omega}_1+\boldsymbol{\omega}_2}{2}\Big)\Big]^2-m\frac{1-\al^2}{4}(\boldsymbol{\widehat{\sigma}}\cdot \boldsymbol{g})^2.
\eeqa
\end{widetext}
The right hand side of Eq.\ \eqref{2.14.1} vanishes (and so, the total energy is conserved in a collision) when $\al=1$ and $\beta=-1$ (perfectly smooth spheres) and $\al=1$ and $\beta=1$ (perfectly rough spheres).

As in our previous works on granular suspensions, \cite{HTG17,HTG20,HT19,GGG19,GGKG20} the effect of the interstitial gas on the inelastic rough hard spheres is accounted for by the operator $\mathcal{F}$ acting on the velocity distribution function $f$. In the case that the spheres are perfectly smooth (and so, inelasticity only affects the translational degrees of freedom of the spheres), for low Reynolds numbers, the instantaneous fluid force is usually constituted by two terms: (i) a drag force term proportional to the relative velocity $\mathbf{v}-\mathbf{U}_g$ ($\mathbf{U}_g$ being the known mean flow velocity of the background gas) and (ii) a stochastic Langevin-like term modeled as a Gaussian white noise. \cite{K81} While the first term (Stokes' law) takes into account the dissipation of energy due to the friction of grains on the viscous gas, the stochastic force gives energy to the solid particles in a random way. This latter term mimics the interaction between the solid particles and the particles of the surrounding (bath) gas. Both terms account for the coupling between the translational degrees of freedom of the spheres and the background gas. Needless to say, one might expect similar effects with the rotational degrees of freedom of grains in the case of inelastic rough spheres.

Therefore, following a generalized Fokker--Planck equation for rotating spheres proposed many years ago by Hess, \cite{H68} we write the operator $\mathcal{F} f$ as
\beq
\label{2.6}
\mathcal{F} f=\mathcal{F}^{\text{tr}} f+\mathcal{F}^{\text{rot}} f,
\eeq
where $\mathcal{F}^{\text{tr}}$ and $\mathcal{F}^{\text{rot}}$ denote the corresponding Fokker--Planck terms associated with the translational and rotational degrees of freedom of spheres. As usual, the translational part $\mathcal{F}^{\text{tr}} f$ can be written as \cite{HTG17,HTG20,GGG19}
\beq
\label{2.7}
\mathcal{F}^{\text{tr}} f=-\gamma_t \frac{\partial}{\partial \mathbf{v}}\cdot \left(\mathbf{v}-\mathbf{U}_g\right)f-\gamma_t \frac{T_\text{ex}}{m} \frac{\partial^2 f}{\partial v^2},
\eeq
where $\gamma_t$ is a drag coefficient associated with the translational degrees of freedom and $T_\text{ex}$ is temperature of the interstitial molecular gas. Although $\gamma_t$ is in general a tensor, it may be considered as a scalar proportional to the viscosity of the background fluid $\eta_g\propto \sqrt{T_\text{ex}}$ in the case of very dilute suspensions. More specifically, if the diameter of the sphere is very large compared with the mean free path of the viscous gas, then $\gamma_t=3\pi \sigma \eta_g/m$. It must be noted that the strength of the correlation in the stochastic term of Eq.\ \eqref{2.7} has been chosen to be consistent with the fluctuation-dissipation theorem when collisions are elastic. \cite{K81} Similarly, the rotational part $\mathcal{F}^{\text{rot}} f$ has an analogous structure to Eq.\ \eqref{2.7} except that the linear velocity $\mathbf{v}$ is replaced by the angular velocity $\boldsymbol{\omega}$. It is given by \cite{H68}
\beq
\label{2.8}
\mathcal{F}^{\text{rot}} f=-\gamma_r \frac{\partial}{\partial \boldsymbol{\omega}}\cdot \boldsymbol{\omega} f-\gamma_r \frac{T_\text{ex}}{m} \frac{\partial^2 f}{\partial \omega^2},
\eeq
where $\gamma_r=\pi \sigma^3 \eta_g/I$. Note that in contrast to $\mathcal{F}^{\text{tr}}$, the ``drag'' term of $\mathcal{F}^{\text{rot}}$ is proportional to the (instantaneous) angular velocity $\boldsymbol{\omega}$; we are assuming for simplicity that the mean angular velocity of the surrounding gas is zero. Moreover, in Eqs.\ \eqref{2.7}--\eqref{2.8}, we are also neglecting a term which takes into account the coupling of translational and rotational motions. This term stems from the transverse force $\mathbf{v} \times \boldsymbol{\omega}$ and was originally proposed in the Brownian model of rotating particles. \cite{H68} \vicente{A consequence of this decoupling is that the solution to the Boltzmann equation from Grad's method \cite{G49} in the uniform shear flow problem is defined in terms of a \emph{two-temperature} Maxwellian distribution [see Eqs.\ \eqref{4.2} and \eqref{4.4}] where the translational and rotational degrees of freedom are not correlated. By using this simple approach, the corresponding contributions to the stress tensor coming from the above transverse force term vanish by symmetry}.
A simpler version of the generalized Fokker--Planck model \eqref{2.6} has been recently employed to study colloidal Brazil nut effect in microswimmer mixtures. \cite{JLHBL19}

According to Eqs.\ \eqref{2.7} and \eqref{2.8}, the Boltzmann kinetic equation \eqref{2.1} can be written as
\begin{widetext}
\beq
\label{2.9}
\frac{\partial f}{\partial t}+\mathbf{v}\cdot \nabla f-\gamma_t\Delta \mathbf{U}\cdot \frac{\partial f}{\partial \mathbf{v}}-\gamma_t\frac{\partial}{\partial \mathbf{v}}\cdot \mathbf{V} f-\gamma_t \frac{T_{\text{ex}}}{m}\frac{\partial^2 f}{\partial v^2}
-\gamma_r \frac{\partial}{\partial \boldsymbol{\omega}}\cdot \boldsymbol{\omega} f-\gamma_r \frac{T_{\text{ex}}}{I}\frac{\partial^2 f}{\partial \omega^2}=J[f,f].
\eeq
\end{widetext}
Here, $\Delta \mathbf{U}=\mathbf{U}-\mathbf{U}_g$,
\beq
\label{2.10}
\mathbf{U}(\mathbf{r};t)=\frac{1}{n(\mathbf{r};t)}\int \dd \mathbf{v} \int \dd\boldsymbol{\omega}\; \mathbf{v}\; f(\mathbf{r}, \mathbf{v}, \boldsymbol{\omega}; t)
\eeq
is the mean flow velocity of spheres, $\mathbf{V}=\mathbf{v}-\mathbf{U}$ is the translational peculiar velocity, and
\beq
\label{2.11}
n(\mathbf{r};t)=\int \dd \mathbf{v} \int \dd\boldsymbol{\omega}\; f(\mathbf{r}, \mathbf{v}, \boldsymbol{\omega}; t)
\eeq
is the number density.

It is quite apparent that the collision dynamics of the suspension model \eqref{2.9} is not affected by the presence of the background gas (namely, the form of the Boltzmann collision operator is the same as that of a \emph{dry} granular gas), and hence we neglect the inertia of the gas phase. As has been widely discussed in several papers on suspensions, \cite{K90,TK95,SMTK96,KH01,WKL03} the above approximation requires that the stresses exerted by the molecular gas on the inelastic rough spheres are sufficiently small to assume that they have a mild impact on the motion of grains. As the particle density decreases with respect to the gas/fluid density (for instance, glass beads in liquid water), the inertia of gas phase is not negligible and hence, the presence of the background gas must be considered in the Boltzmann collision operator.

\subsection{Balance equations}

The transfer equation for an arbitrary dynamic property $\psi(\mathbf{r}, \mathbf{v}, \boldsymbol{\omega}, t)$ can be obtained by multiplying both sides of the Boltzmann equation \eqref{2.9} by $\psi$ and integrating over $\mathbf{v}$ and $\boldsymbol{\omega}$. In order to obtain the transfer equation, an useful property of the Boltzmann collision operator is \cite{G19}
\beqa
\label{2.12}
\mathcal{J}[\psi|f,f]&\equiv&\int \dd{\bf v}_1\int \dd\boldsymbol{\omega}_1  \psi(\mathbf{r}, \mathbf{v}_1, \boldsymbol{\omega}_1)J[\mathbf{v}_1,\boldsymbol{\omega}_1|f,f]\nonumber\\
&=&\sigma^2 \int \dd{\bf v}_1\int \dd\boldsymbol{\omega}_1 \int \dd{\bf v}_2 \int \dd\boldsymbol{\omega}_2\int \dd\widehat{\boldsymbol{\sigma}}\,\Theta (\widehat{{\boldsymbol {\sigma}}}\cdot {\bf g})\nonumber\\
& & \times (\widehat{\boldsymbol {\sigma }}\cdot {\bf g})\Big[\psi(\mathbf{r},\mathbf{v}_1', \boldsymbol{\omega}_1')-
\psi(\mathbf{r},\mathbf{v}_1, \boldsymbol{\omega}_1)\Big],
\eeqa
where the collisional rules for the direct collision are given by Eqs.\ \eqref{2.13} and \eqref{2.14}.

The evolution equation for the average
\beq
\label{2.15}
\langle \psi \rangle =\frac{1}{n(\mathbf{r},t)}\int \dd{\bf v}\int \dd\boldsymbol{\omega} \; \psi(\mathbf{r},\mathbf{v}, \boldsymbol{\omega};t)f(\mathbf{r},\mathbf{v}, \boldsymbol{\omega};t)
\eeq
can be now easily obtained with the result
\beqa
\label{2.16}
& & \frac{\partial}{\partial t} (n \langle \psi \rangle )-n\langle \frac{\partial \psi}{\partial t} \rangle+\nabla \cdot \left(n \langle \mathbf{v} \psi \rangle\right)
-n\langle \mathbf{v}\cdot \nabla \psi \rangle \nonumber\\
& & +n\gamma_t \Delta \mathbf{U}\cdot \langle \frac{\partial \psi}{\partial \mathbf{v}} \rangle
+n \gamma_t \langle \mathbf{V}\cdot \frac{\partial \psi}{\partial \mathbf{v}} \rangle-n\frac{\gamma_t T_\text{ex}}{m} \langle \frac{\partial^2\psi}{\partial v^2} \rangle \nonumber\\
& & +n \gamma_r \langle \boldsymbol{\omega}\cdot \frac{\partial \psi}{\partial \boldsymbol{\omega}} \rangle-n\frac{\gamma_r T_\text{ex}}{I} \langle \frac{\partial^2\psi}{\partial \omega^2} \rangle=\mathcal{J}[\psi|f,f].
\eeqa
The macroscopic balance equations for the densities of mass, momentum, and energy can be obtained from the transfer equation \eqref{2.16} when $\psi\equiv \left\{1, m\mathbf{v}, mV^2/2+I\omega^2/2\right\}$. They are given by
\beq
\label{2.17}
D_t n+n\nabla \cdot \mathbf{U}=0,
\eeq
\beq
\label{2.18}
\rho D_t \mathbf{U}=-\rho \gamma_t \Delta \mathbf{U}-\nabla \cdot \mathsf{P},
\eeq
\beqa
\label{2.19}
& & D_t T+\gamma_t\left(T_t-T_\text{ex}\right)+\gamma_r \left(T_r-T_\text{ex}\right)=-\zeta T
\nonumber\\
& &
-\frac{1}{3n}\left(\nabla \cdot \mathbf{q}+\mathsf{P}:\nabla \mathbf{U}\right).
\eeqa
In Eqs.\ \eqref{2.17}--\eqref{2.19}, $\rho=m n$ is the mass density, $D_t\equiv \partial_t+\mathbf{U}\cdot \nabla$ is the material time derivative, and the granular temperature $T(\mathbf{r},t)$ is defined as
\beq
\label{2.20}
T=\frac{1}{2}\left(T_t+T_r\right),
\eeq
where the (partial) translational $T_t$ and rotational $T_r$ temperatures are defined as
\beq
\label{2.21}
T_t=\frac{m}{3}\langle V^2 \rangle, \quad T_r=\frac{I}{3}\langle \omega^2 \rangle,
\eeq
where the averages $\langle \cdots \rangle$ are defined by Eq.\ \eqref{2.15}. Moreover, the pressure tensor $\mathsf{P}(\mathbf{r},t)$ is
\beq
\label{2.20.1}
\mathsf{P}=\rho \langle \mathbf{V}\mathbf{V} \rangle,
\eeq
while the heat flux vector $\mathbf{q}(\mathbf{r},t)$ is given by
\beq
\label{2.21.1}
\mathbf{q}=\mathbf{q}_t+\mathbf{q}_r,
\eeq
where the translational $\mathbf{q}_t$ and rotational $\mathbf{q}_r$ contributions are defined as
\beq
\label{2.22}
\mathbf{q}_t=\frac{\rho}{2}\langle V^2 \mathbf{V}\rangle, \quad
\mathbf{q}_r=\frac{In}{2}\langle \omega^2 \mathbf{V}\rangle.
\eeq
Moreover, the cooling rate $\zeta$ (which gives the rate of energy dissipation due to inelasticity) is
\beq
\label{2.23}
\zeta=\frac{T_t}{2T}\zeta_t+\frac{T_r}{2T}\zeta_r,
\eeq
where the partial energy production rates associated with the translational ($\zeta_t$) and rotational ($\zeta_r$) degrees of freedom are
\beq
\label{2.24}
\zeta_t=-\frac{m}{3nT_t}\mathcal{J}[v^2|f,f], \quad
\zeta_r=-\frac{I}{3nT_r}\mathcal{J}[\omega^2|f,f].
\eeq
One third of the trace of the pressure tensor $\mathsf{P}$ defines the hydrostatic pressure $p$ as
\beq
\label{2.25}
p=n T_t.
\eeq

At a kinetic theory level, it is also convenient to derive the balance equations for the partial temperatures $T_t$ and $T_r$. They are given by
\beq
\label{2.26}
D_t T_t+2\gamma_t\left(T_t-T_\text{ex}\right)+\zeta_t T_t=-\frac{2}{3n}\left(\nabla \cdot \mathbf{q}_t+\mathsf{P}:\nabla \mathbf{U}\right),
\eeq
\beq
\label{2.27}
D_t T_r+2\gamma_r\left(T_r-T_\text{ex}\right)+\zeta_r T_r=-\frac{2}{3n}\nabla \cdot \mathbf{q}_r.
\eeq
Combination of Eqs.\ \eqref{2.26} and \eqref{2.27} leads to Eq.\ \eqref{2.19}.

Before finishing this section, it is worthwhile remarking that in the definition of $T_r$ [second relation of Eq.\ \eqref{2.21}] we have not referred the angular velocities $\boldsymbol{\omega}$ to the mean value $\boldsymbol{\Omega}=\langle \boldsymbol{\omega} \rangle$. This contrasts with the definition of $T_t$ [first relation of Eq.\ \eqref{2.21}] where the (instantaneous) velocity $\mathbf{v}$ has been referred to $\mathbf{U}$. As noted in previous works, \cite{SKG10} we have not defined $T_r$ in terms of the difference $\boldsymbol{\omega}-\boldsymbol{\Omega}$ because $\boldsymbol{\Omega}$ is not a conserved quantity. In the case that we were defined the rotational temperature as $\widetilde{T}_r=\frac{I}{3}\langle \left(\boldsymbol{\omega}-\boldsymbol{\Omega}\right)^2 \rangle$, then the granular temperature $\widetilde{T}=(\widetilde{T}_t+\widetilde{T}_r)/2$ would not be a conserved hydrodynamic field in the case of elastic ($\al=1$) and completely rough ($\beta=1$) spheres, even although the total energy is conserved in collisions [see Eq.\ \eqref{2.14.1} where $\Delta E=0$ if $\al=\beta=1$].

\section{Simple shear flow}
\label{sec3}

We assume that the inertial suspension is under simple (uniform) shear flow. As described in many previous works, \cite{G19} this state is macroscopically characterized by a constant number density $n$, a uniform granular temperature $T(t)$, and macroscopic velocity
field
\beq
\label{3.1}
U_i=a_{ij}r_j, \quad a_{ij}=a\delta_{ix}\delta_{jy},
\eeq
$a$ being the \emph{constant} shear rate. We also assume that the mean angular velocity $\boldsymbol{\Omega} =\mathbf{0}$ and, as usual in uniform sheared suspensions, the average (linear) velocity of particles follows the velocity of the fluid phase: $\mathbf{U}=\mathbf{U}_g$.
At a microscopic level, the main advantage of the simple shear flow is that this state becomes spatially homogeneous when the velocities of the particles $\mathbf{v}$ are referred to the frame moving with the linear velocity field $\mathbf{U}$. \cite{DSBR86,GS03} In this frame, the distribution function has the form  $f(\mathbf{r}, \mathbf{v}, \boldsymbol{\omega}; t)=f(\mathbf{V}, \boldsymbol{\omega}; t)$ and hence, the Boltzmann equation \eqref{2.9} becomes
\beqa
\label{3.2}
& & \frac{\partial f}{\partial t}-aV_y\frac{\partial f}{\partial V_x}-\gamma_t\frac{\partial}{\partial \mathbf{v}}\cdot \mathbf{V} f-\gamma_t \frac{T_{\text{ex}}}{m}\frac{\partial^2 f}{\partial v^2}-\gamma_r\frac{\partial}{\partial \boldsymbol{\omega}}\cdot \boldsymbol{\omega} f\nonumber\\
& & -\gamma_r \frac{T_{\text{ex}}}{I}\frac{\partial^2 f}{\partial \omega^2}=J[f,f].
\eeqa

Since $\nabla n=\nabla T=0$, the heat flux vanishes ($\mathbf{q}=\mathbf{0}$) in the simple shear flow and the (uniform) pressure tensor $\mathsf{P}$ is the relevant irreversible flux of the problem. The knowledge of $\mathsf{P}$ allows us to identify the most significant non-Newtonian transport properties of the suspension.

In the simple shear flow problem, the conservation equations \eqref{2.17} and \eqref{2.18} applies trivially while the balance equations \eqref{2.26} and \eqref{2.27} for the translational $T_t$ and rotational $T_r$ temperatures, respectively, yield
\beq
\label{3.3}
\frac{\partial T_t}{\partial t}+2\gamma_t\left(T_t-T_\text{ex}\right)+\zeta_t T_t=-\frac{2a}{3n}P_{xy},
\eeq
\beq
\label{3.4}
\frac{\partial T_r}{\partial t}+2\gamma_r\left(T_r-T_\text{ex}\right)+\zeta_r T_r=0.
\eeq
Note that the (partial) energy production rates $\zeta_t$ and $\zeta_r$ are defined in terms of the velocity distribution function $f(\mathbf{V}, \boldsymbol{\omega})$ [see Eqs.\ \eqref{2.24}]. This means that one has necessarily to get a solution of the Boltzmann equation \eqref{3.2} to determine $\zeta_t$ and $\zeta_r$ and the stress tensor $P_{xy}$. Once the above quantities are known, then the partial temperatures $T_t$ and $T_r$ can be obtained by solving Eqs.\ \eqref{3.3} and \eqref{3.4}.

According to Eqs.\ \eqref{3.3} and \eqref{3.4}, there two competing mechanisms in the time evolution of the temperature. On the one hand, there are cooling terms arising from inelastic cooling and the friction of grains on viscous gas. On the other hand, there are heating terms arising from the viscous heating and the energy provided to the particles by the stochastic driving term. After a transient period, one expects that both mechanisms compensate for each other and a \emph{steady} state is achieved.

In the absence of shear rate ($a=0$) and in the steady state ($\partial_t f=0$), for $\al=1$ and $|\beta|=1$ the total kinetic energy is conserved, and the solution to Eq.\ \eqref{3.2} is given by the Maxwellian velocity distribution
\beq
\label{3.5}
f_\text{M}(\mathbf{V}, \boldsymbol{\omega})=n\Big(\frac{m I}{4\pi^2 T_\text{ex}^2}\Big)^{3/2} \exp\left(-\frac{mv^2}{2T_\text{ex}}\right)
\exp\left(-\frac{I\omega^2}{2T_\text{ex}}\right).
\eeq
On the other hand, beyond the above two special cases, the solution to Eq.\ \eqref{3.2} is not known.

The relevant elements of the pressure tensor may be obtained by multiplying both sides of Eq.\ \eqref{3.2} by $m V_k V_\ell$ and integrating over $\mathbf{V}$ and $\boldsymbol{\omega}$. The result is
\beqa
\label{4.1}
\partial_t P_{k\ell}+a_{kj}P_{\ell j}+a_{j\ell}P_{jk}&+&2\gamma_t \left(P_{k\ell}-nT_\text{ex}\delta_{k\ell}\right)\nonumber\\
&=& m\mathcal{J}[V_kV_\ell|f,f].
\eeqa
On the other hand, the exact form of $\mathcal{J}[V_kV_\ell|f,f]$ is not known, even in the simplest case $\al=1$ and $\beta=\pm 1$ where the kinetic energy is conserved in collisions. Thus, one has to resort to alternative approaches for computing the pressure tensor $P_{ij}$. As mentioned in the Introduction, in this paper we will determine the elements of the pressure tensor by using two different but complementary routes: (i) by solving the Boltzmann equation by means of Grad's moment method, and (ii) by considering a BGK-like kinetic model recently proposed \cite{S11} for inelastic rough hard spheres.


\section{Grad's moment method}
\label{sec4}

As has been clearly shown in several previous works, \cite{G13,ChVG15,HTG17,HTG20} Grad's moment method can be considered as an accurate tool to estimate the collisional moment $\mathcal{J}[V_kV_\ell|f,f]$. In the same way as in molecular fluids, \cite{G49} the idea of Grad's method is to expand the velocity distribution function in powers of generalized Hermite polynomials, the coefficients of the expansion being the corresponding velocity moments. This expansion is truncated at a given order $k$ and so, the moments of degree higher than $k$ are neglected in the corresponding solution. In the case of a three-dimensional gas, the usual thirteen-moment approximation includes the density $n$, the three components of the mean flow velocity $\mathbf{U}$, the six elements of the pressure tensor $\mathsf{P}$ [recall that $T_t=(1/3n)(P_{xx}+P_{yy}+P_{zz}$)], and the three components of the heat flux vector $\mathbf{q}$. \cite{G49,S05} Since the heat flux vanishes in the simple shear flow problem, then Grad's solution is given by\cite{L91,DT75}
\beq
\label{4.2}
f(\mathbf{V}, \boldsymbol{\omega})\to f_0(\mathbf{V}, \boldsymbol{\omega})\Big[1+\frac{m}{2n T_t^2}\left(V_i V_j-\frac{1}{3}V^2 \delta_{ij}\right)\Pi_{ij}\Big],
\eeq
where
\beq
\label{4.3}
\Pi_{ij}=P_{ij}-p\delta_{ij}
\eeq
is the traceless part of the pressure tensor and $f_0$ is the \emph{two-temperature} Maxwellian velocity distribution
\beq
\label{4.4}
f_0(\mathbf{V}, \boldsymbol{\omega})=n\Big(\frac{m I}{4\pi^2 T_t T_r}\Big)^{3/2} \exp\left(-\frac{mV^2}{2T_t}\right)
\exp\left(-\frac{I\omega^2}{2T_r}\right).
\eeq
Upon writing the distribution \eqref{4.2} we have ignored the possible contributions to $f$ coming from the combination of traceless dyadic products of the three vectors $\mathbf{V}$, $\left(\mathbf{V}\cdot \boldsymbol{\omega}\right)$, and $\mathbf{V}\times \boldsymbol{\omega}$ with unknown scalar coefficients.\cite{KSG14} These contributions are absent because we have neglected the orientational correlations between $\mathbf{V}$ and $\boldsymbol{\omega}$ \vicente{in the Fokker--Planck operator $\mathcal{F}$ [see Eqs.\ \eqref{2.6}--\eqref{2.8}]. Thanks to this simplification, we resort to} the weight distribution $f_0$, which is isotropic in velocity space. In addition, we have also neglected in Grad's solution \eqref{4.4} the contribution of the fourth-degree velocity moments (cumulants) to the distribution $f$. These cumulants have been determined in homogeneous situations, \cite{SKS11,VSK14,VS15} showing that in general these quantities are small, specially when the system is driven by a white-noise stochastic thermostat.\cite{SKS11,VS15} On the other hand, in spite of the above approximations, it is worthwhile noticing that the theoretical predictions for the temperature ratio $T_r/T_t$ obtained by replacing $f$ by $f_0$ in homogeneous states have been shown to compare very well with Monte Carlo and molecular dynamics simulations. \cite{VSK14} We expect that this fair agreement is also kept in the simple shear flow state.

The collisional moment $\mathcal{J}[V_kV_\ell|f,f]$ can be computed when the trial distribution \eqref{4.2} is inserted into the definition of this moment. The calculations are long but standard and are based on the relationship \eqref{2.14}. After some algebra, one gets \cite{L91,KSG14}
\beq
\label{4.5}
m\mathcal{J}[V_k V_\ell|f,f]=-\nu_\eta \Pi_{k\ell}-p \zeta_t \delta_{k\ell},
\eeq
where we recall that $p=n T_t$, and
\beq
\label{4.6}
\nu_\eta=\Bigg[\left(\widetilde{\alpha}+\widetilde{\beta}\right)\left(2-\widetilde{\al}-\widetilde{\beta}\right)+
\frac{\widetilde{\beta}^2}{6\kappa}\frac{T_r}{T_t}\Bigg]\nu_t,
\eeq
\beq
\label{4.7}
\zeta_t=\frac{5}{3}\Bigg[\widetilde{\alpha}(1-\widetilde{\alpha})+\widetilde{\beta}(1-\widetilde{\beta})-
\frac{\widetilde{\beta}^2}{\kappa}\frac{T_r}{T_t}\Bigg]\nu_t.
\eeq
In Eqs.\ \eqref{4.6}--\eqref{4.7},
\beq
\label{4.8}
\widetilde{\alpha}=\frac{1+\al}{2}, \quad \widetilde{\beta}=\frac{\kappa}{1+\kappa}\frac{1+\beta}{2},
\eeq
and $\nu_t$ is the effective collision frequency
\beq
\label{4.9}
\nu_t=\frac{16}{5}n\sigma^2\sqrt{\frac{\pi T_t}{m}}.
\eeq
In addition, the cooling rate $\zeta_r$ associated with the rotational degrees of freedom [defined by the second relation of Eq.\ \eqref{2.24}] can be also determined from the Grad's distribution \eqref{4.2} with the result \cite{L91,KSG14}
\beq
\label{4.10}
\zeta_r=\frac{5}{6}\frac{\widetilde{\beta}}{\kappa}\Bigg[1-\beta+2\widetilde{\beta}
\left(1-\frac{T_t}{T_r}\right)\Bigg]\nu_t.
\eeq
Upon deriving Eq.\ \eqref{4.5}, nonlinear terms in the tensor $\Pi_{k\ell}$ have been neglected. Equation \eqref{4.1} can be more explicitly written when the expression \eqref{4.5} is accounted for. The result is
\beqa
\label{4.11}
\partial_t P_{k\ell}+a_{kj}P_{j\ell}+a_{\ell j}P_{jk}&+&2\gamma_t \left(P_{k\ell}-nT_\text{ex}\delta_{k\ell}\right)=
-\nu_\eta P_{k \ell}\nonumber\\
& & -p\left(\zeta_t-\nu_\eta\right)\delta_{k\ell}.
\eeqa
Equation \eqref{4.11} clearly shows that $P_{yy}=P_{zz}$ and hence, the constraint \eqref{2.25} yields $P_{xx}=3p-2 P_{yy}$.
The equality $P_{yy}=P_{zz}$ do not agree with computer simulation results obtained for smooth granular suspensions. \cite{TK95,ChVG15}
The above drawback could be fixed if one would retain nonlinear terms in $\Pi_{k\ell}$ in the evaluation of $\mathcal{J}[V_kV_\ell|f,f]$. The inclusion of these nonlinear
corrections provides nonzero contributions to the normal stress differences in the plane orthogonal to the shear flow
(namely, $P_{yy}-P_{zz}\neq 0$). \cite{ChVG15} However, the difference $P_{yy}-P_{zz}$ is in general very small and so, the expression \eqref{4.5} can be still considered as a good approximation.

It is convenient now to introduce dimensionless quantities. Among the different possibilities, as in previous works on sheared granular suspensions, \cite{HTG17,HTG20,GGG19} we scale the quantities associated with the solid particles with those referring to the gas phase, namely, $\gamma_t$, $\gamma_r$, and $T_\text{ex}$. Since the pressure tensor (which is the most relevant flux in the simple shear flow state) is mainly related to the translational degrees of freedom, we reduce here the shear rate and the external temperature with respect to the (translational) friction coefficient $\gamma_t$, namely,
\beq
\label{4.14}
a^*\equiv \frac{a}{\gamma_t}, \quad T_\text{ex}^*\equiv \frac{T_\text{ex}}{m\sigma^2 \gamma_t^2}.
\eeq
In addition, the translational and rotational temperatures are scaled with respect to $T_\text{ex}$ ($\theta_t\equiv T_t/T_\text{ex}$ and $\theta_r\equiv T_r/T_\text{ex}$) and we introduce the dimensionless quantities
\beqa
\label{4.15}
\zeta_t^*\equiv \frac{\zeta_t}{\sqrt{\theta_t}\gamma_t}
&=&\frac{16}{3}\sqrt{\pi}\Bigg[\widetilde{\alpha}(1-\widetilde{\alpha})+
\widetilde{\beta}(1-\widetilde{\beta})-\frac{\widetilde{\beta}^2}{\kappa}\frac{\theta_r}{\theta_t}\Bigg]\nonumber\\
& & \times n^*\sqrt{T_\text{ex}^*},
\eeqa
\beqa
\label{4.16}
\nu_\eta^*\equiv \frac{\nu_\eta}{\sqrt{\theta_t}\gamma_t}&=&\frac{16}{5}\sqrt{\pi}\Bigg[\left(\widetilde{\alpha}+\widetilde{\beta}\right)
\left(2-\widetilde{\al}-\widetilde{\beta}\right)+\frac{\widetilde{\beta}^2}{6\kappa}\frac{\theta_r}{\theta_t}\Bigg]
\nonumber\\
& & \times n^*\sqrt{T_\text{ex}^*}.
\eeqa
Here, $n^*\equiv n\sigma^3$ is the reduced density. As already noted in previous studies,\cite{GGG19} the explicit dependence of $\zeta_t^*$ and $\nu_\eta^*$ on density comes from the dimensionless quantities $a^*$ and $T_\text{ex}^*$. This way of reducing the above quantities is closer to the one made in computer simulations for smooth inelastic hard spheres. \cite{HTG17,HTG20} Needless to say, if you had reduced $a$ and $T_\text{ex}$ with the collision frequency $\nu_t$ (this sort of scaling is usual in sheared molecular gases\cite{EM90}), the above density dependence had been removed. Note that $\zeta_t^*$ and $\nu_\eta^*$ are independent of both the (translational) temperature $T_t$ and the background temperature $T_\text{ex}$ because $\gamma_t \propto \sqrt{T_\text{ex}}$.

In terms of the above dimensionless variables, the set of Eqs.\ \eqref{4.11} become
\beqa
\label{4.16.1}
\partial_\tau P_{k\ell}^*+a_{kj}^*P_{j\ell}^*+a_{\ell j}^*P_{jk}^*&+&2\left(P_{k\ell}^*-\delta_{k\ell}\right)=
-\nu_\eta^*\sqrt{\theta_t} P_{k \ell}^*\nonumber\\
& & -\theta_t\sqrt{\theta_t}\left(\zeta_t^*-\nu_\eta^*\right)\delta_{k\ell},
\eeqa
where we have introduced the (scaled) time variable $\tau$ defined as $\dd \tau=\gamma_t \dd t$.

\subsection{Steady state solution}

As said before, after a transient regime, one expects that the suspension reaches a \emph{steady} state. The interesting point is that this steady sheared state is inherently non-Newtonian. \cite{SGD04} The main goal of this paper is to determine the rheological properties of the inertial suspension in the steady uniform shear flow.

An inspection to the results derived in the smooth case \cite{GGG19} shows that Eq.\ \eqref{4.16.1} (with $\partial_\tau P_{k\ell}^*=0$) is formally equivalent to that of this limit case when one makes the changes $\theta\to \theta_t$, $\zeta^*\to \zeta_t^*$, and $\nu_{0|2}^*\to \nu_\eta^*$, where the quantities $\theta$, $\zeta^*$, and $\nu_{0|2}^*$ are defined in Ref.\ \onlinecite{GGG19}. Consequently, the expressions of $P_{yy}^*$, $P_{xy}^*$, and $a^*$ can be obtained from comparison with those obtained in the smooth case [see Eqs.\ (32), (33), and (35) of Ref.\ \onlinecite{GGG19}]. They are given by
\beq
\label{4.17}
P_{yy}^*=P_{zz}^*=\frac{2+\left(\nu_\eta^*-\zeta_t^*\right)\theta_t\sqrt{\theta_t}}{2+\sqrt{\theta_t}\nu_\eta^*}, \quad P_{xx}^*=3  \theta_t-2P_{yy}^*,
\eeq
\beq
\label{4.18}
P_{xy}^*=-\frac{2+\left(\nu_\eta^*-\zeta_t^*\right)\theta_t\sqrt{\theta_t}}{(2+\sqrt{\theta_t}\nu_\eta^*)^2}a^*.
\eeq
\beq
\label{4.19}
a^*=\sqrt{\frac{3}{2}\frac{\sqrt{\theta_t}\zeta_t^*+2\left(1-\theta_t^{-1}\right)}{\sqrt{\theta_t}\left(\nu_\eta^*-\zeta_t^*\right)
+2\theta_t^{-1}}}\Big(2+\sqrt{\theta_t}\nu_\eta^*\Big),
\eeq
The steady (reduced) temperatures $\theta_t$ and $\theta_r$ can be determined from Eqs.\ \eqref{3.3} and \eqref{3.4} (with $\partial_t\theta_t=\partial_t \theta_r=0$) as
\beq
\label{4.19.1}
2\left(\theta_t-1\right)+\sqrt{\theta_t}\theta_t \zeta_t^*=-\frac{2}{3}a^* P_{xy}^*,
\eeq
\beq
\label{4.20}
2\frac{\gamma_r}{\gamma_t}\left(\theta_r-1\right)+ \sqrt{\theta_t}\theta_r \zeta_r^*=0,
\eeq
where $\gamma_r/\gamma_t=4/(3\kappa)$ and
\beq
\label{4.20.1}
\zeta_r^*\equiv \frac{\zeta_r}{\sqrt{\theta_t}\gamma_t}
=\frac{8}{3}\sqrt{\pi}\frac{\widetilde{\beta}}{\kappa}\Bigg[1-\beta+2\widetilde{\beta}
\left(1-\frac{\theta_t}{\theta_r}\right)\Bigg]n^*\sqrt{T_\text{ex}^*}.
\eeq
On the other hand, as already happens in smooth granular suspensions, \cite{HTG17,HTG20,GGG19} it is not possible to express in Eq.\ \eqref{4.19} $\theta_t$ in terms of $a^*$ and the remaining parameters of the suspension. Thus, for given values of $\al$, $\beta$, $\kappa$, $n^*$, and $T_\text{ex}^*$, one can consider $\theta_t$ for instance as input parameter and determine $a^*$ and $\theta_r$ as the solutions to Eqs.\ \eqref{4.19} and \eqref{4.20}.

Once the (scaled) translational temperature $\theta_t$ is known, the rheological properties of the suspension are obtained from Eqs.\ \eqref{4.17}, \eqref{4.18}, and \eqref{4.20}. In particular, the (dimensionless) non-Newtonian shear viscosity
\beq
\label{4.20.2}
\eta^*=\frac{P_{xy}^*}{a^*}
\eeq
is given by
\beq
\label{4.21}
\eta^*=\frac{2+\left(\nu_\eta^*-\zeta_t^*\right)\sqrt{\theta_t}\theta_t}{\left(2+\sqrt{\theta_t}\nu_\eta^*\right)^2}.
\eeq
Since (linear) Grad's solution \eqref{4.16.1} yields $P_{yy}^*=P_{xx}^*$, then the only nonvanishing viscometric function is the one associated with the difference $P_{xx}^*-P_{yy}^*$. In dimensionless form, the first viscometric function is defined as
\beq
\label{4.22}
\Psi^*=P_{xx}^*-P_{yy}^*=3\theta_t\frac{2\left(1-\theta_t^{-1}\right)+\sqrt{\theta_t}\zeta_t^*}{2+\sqrt{\theta_t}\nu_\eta^*}.
\eeq
As expected, the expressions \eqref{4.19}, \eqref{4.21}, and \eqref{4.22} agree with the ones derived for inelastic Maxwell models \cite{GGG19}  [see Eqs.\ (35), (39), and (40) of Ref.\ \onlinecite{GGG19}] when one makes the replacements $\theta_t\to \theta$, $\zeta_t^*\to \zeta^*$, and $\nu_\eta^*\to \nu_{0|2}^*$, where the quantities $\theta$, $\zeta^*$, and $\nu_{0|2}^*$.

\subsection{Navier--Stokes results}

In order to get analytical results, it is illustrative to consider the limits of small and large shear rates. First, when $a^*\to 0$, Eq.\ \eqref{4.19} yields the following relation for determining the (translational) temperature $\theta_t^{(0)}$:
\beq
\label{4.23}
\theta_t^{(0)}\Big(1+\frac{1}{2}\zeta_t^*\sqrt{\theta_t^{(0)}}\Big)-1=0.
\eeq
The rotational temperature $\theta_r^{(0)}$ is easily obtained from Eq.\ \eqref{4.20} as
\beq
\label{4.24}
\theta_r^{(0)}=\Big(1+\frac{1}{2}\frac{\gamma_t}{\gamma_r}\zeta_r^*\sqrt{\theta_t^{(0)}}\Big)^{-1}.
\eeq
Substitution of Eq.\ \eqref{4.23} into Eq.\ \eqref{4.21} gives the form of the Navier--Stokes shear viscosity $\eta_\text{NS}^*$:
\beq
\label{4.24}
\eta_\text{NS}^*=\frac{\theta_t^{(0)}}{2+\sqrt{\theta_t^{(0)}}\nu_\eta^*}.
\eeq

In the opposite limit ($a^*\to \infty$), the asymptotic expressions for $\zeta_t^*\neq 0$, $\al<1$, and $|\beta|\neq 1$ are
\beq
\label{4.25}
\theta_t^{(\infty)}=\frac{2}{3}\frac{\nu_\eta^*-\zeta_t^*}{\nu_\eta^{*2}\zeta_t^*}a^{*2},
\quad \eta_\infty^*=\sqrt{\frac{3}{2}}\frac{\left(\nu_\eta^*-\zeta_t^*\right)^{3/2}}{\nu_\eta^{*3/2}\sqrt{\zeta_t^*}}a^{*}.
\eeq
When $\zeta_t^*=0$, one has $\theta_t^{(\infty)}=a^{*4}/(9\nu_\eta^{*2})$ and $\eta_{\infty}^*=a^{*2}/(3\nu_\eta^{*2})$. The corresponding expressions for $\theta_r^{(\infty)}$ can be obtained from \eqref{4.24} by replacing $\theta_t^{(0)}$ by $\theta_t^{(\infty)}$.

\begin{figure}
\includegraphics[width=0.8\columnwidth]{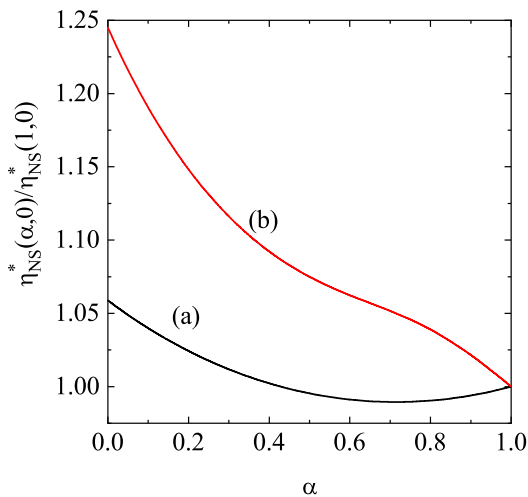}
\caption{Plot of the ratio $\eta_\text{NS}^*(\al)/\eta_\text{NS}^*(1)$ versus the coefficient of normal restitution $\al$ for granular suspensions (a) and dry granular gases (b). Here, we have assumed spheres with a uniform mass distribution ($\kappa=\frac{2}{5}$) and a coefficient of tangential restitution $\beta=1$.}
\label{fig1}
\end{figure}
\begin{figure}
\includegraphics[width=0.8\columnwidth]{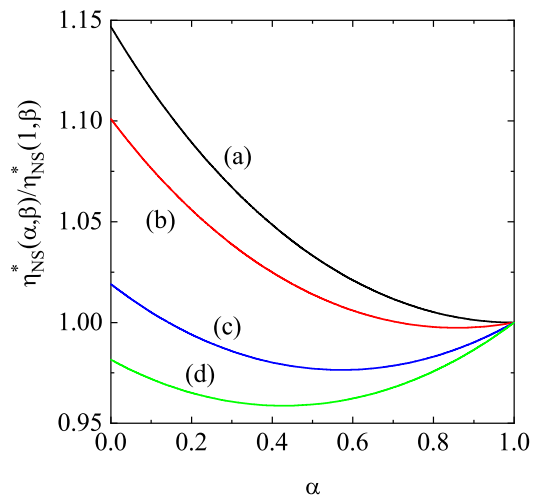}
\caption{Plot of the ratio $\eta_\text{NS}^*(\al,\beta)/\eta_\text{NS}^*(1,\beta)$ versus the coefficient of normal restitution $\al$ for $\kappa=\frac{2}{5}$ and four different values of coefficient of tangential restitution $\beta$: $\beta=-1$ (a), $\beta=-0.5$ (b), $\beta=0.5$ (c), and $\beta=1$ (d). Here, $\eta_\text{NS}^*(1,\beta)$ is given by Eq.\ \eqref{4.24} with $\al=1$.}
\label{fig2}
\end{figure}

Similarly to suspensions of smooth inelastic hard spheres, \cite{HTG17,HTG20,GGG19} Eqs.\ \eqref{4.24} and \eqref{4.25} clearly show that while $\eta^*$ is finite in the Navier--Stokes domain, it diverges for very large shear rates. In fact, the ratio $\eta^*(a^*\to \infty)/\eta^*(a^*\to 0)$ becomes very large as the shear rate increase; this could explain the existence of DST of the shear viscosity coefficient. As mentioned in Sec.\ \ref{sec1}, this behavior gradually changes as the density increases since the theoretical results derived from the Enskog kinetic theory (and confirmed by molecular dynamics simulations) show CST for finite densities. \cite{HTG17,HTG20}

Although we are mainly in this paper interested in non-Newtonian transport properties, Eq.\ \eqref{4.24} gives the expression of the Navier--Stokes shear viscosity coefficient of a suspension of inelastic rough hard spheres. We are not aware of any previous derivation of this relevant transport coefficient. On the other hand, in the absence of the interstitial gas (\emph{dry} granular gas), the Navier--Stokes shear viscosity coefficient was obtained in Ref.\ \onlinecite{KSG14}. Its explicit form is provided in the Appendix \ref{appB} for the sake of completeness. It is quite apparent that the form of the Navier--Stokes shear viscosity of a dry gas of inelastic rough hard spheres [see Eq.\ \eqref{b1}] differs from the one derived here [see Eq.\ \eqref{4.24}], as expected. To illustrate these differences with and without interstitial gas, Fig.\ \ref{fig1} shows the $\al$-dependence of the ratios $\eta_\text{NS}^*(\al)/\eta_\text{NS}^*(1)$ for granular suspensions (line (a)) and dry granular gases (line (b)). In the dry case, $\eta_\text{NS}^*=\eta_\text{NS} \nu_t/(n T_t)$. In Fig.\ \ref{fig1}, $\kappa=\frac{2}{5}$, $\beta=0$, and $\eta_\text{NS}^*(1)$ refers to the value of the shear viscosity at $\al=1$. We observe that the dependence of the ratio $\eta_\text{NS}^*(\al)/\eta_\text{NS}^*(1)$ on $\al$ is very different in both systems, even at a qualitative level since while this ratio exhibits a non-monotonic dependence on the coefficient of normal restitution in the case of granular suspensions, it increases with decreasing $\al$ in the dry granular case. Regarding granular suspensions and to show the combined effect of $\al$ and $\beta$ on $\eta_\text{NS}^*$, Fig.\ \ref{fig2} plots the ratio $\eta_\text{NS}^*(\al,\beta)/\eta_\text{NS}^*(1,\beta)$ as a function of $\al$ for different values of $\beta$. We observe that, at fixed $\al$, the above ratios present a monotonic $\beta$-dependence since those coefficients decreasing from $\beta=-1$ to $\beta=1$. In addition, at fixed $\beta$, we see that while those coefficients increase with decreasing $\al$ when $\beta$ is negative, they exhibit a non-monotonic dependence on $\al$ when $\beta$ is positive. In any case, Fig.\ \ref{fig2} highlights the intricate interplay between the coefficients of restitution $\al$ and $\beta$ on the behavior of the Navier--Stokes shear viscosity coefficient.

\section{BGK-like kinetic model of the Boltzmann equation}
\label{sec5}

To complement the results derived from the Boltzmann equation from Grad's moment method, we consider now a BGK-like kinetic model for a granular gas of inelastic rough hard spheres. \cite{S11} As usual in kinetic models, the intricate mathematical structure of the Boltzmann collision operator $J[\mathbf{v}, \boldsymbol{\omega}|f,f]$ is replaced by a simpler term $K[\mathbf{v}, \boldsymbol{\omega}|f]$ that retains the basic physical properties of the true Boltzmann operator. More specifically, $J[f,f]$ is substituted by the sum of three terms:\cite{S11}  (i) a relaxation term towards a two-temperature local equilibrium distribution, (ii) a nonconservative drag force proportional to $\mathbf{V}$, and (iii) a nonconservative torque equal to a linear combination of $\boldsymbol{\omega}$ and $\boldsymbol{\Omega}$. In the context of the simple shear flow problem, the operator $K[\mathbf{v}, \boldsymbol{\omega}|f]$ becomes
\beqa
\label{5.1}
K[\mathbf{v}, \boldsymbol{\omega}|f]&=&-\chi(\al,\beta)\nu_t\left(f-f_0\right)\nonumber\\
& & +\frac{\zeta_t}{2}\frac{\partial}{\partial \mathbf{V}}\cdot \left(\mathbf{V}f\right)+\frac{\zeta_r}{2}\frac{\partial}{\partial \boldsymbol{\omega}}\cdot \left(\boldsymbol{\omega}f\right),
\eeqa
where $\nu_t$ is the collision frequency defined by Eq.\ \eqref{4.9}, $f_0$ is given by Eq.\ \eqref{4.4}, and the forms of $\zeta_t$ and $\zeta_r$ are provided by Eqs.\ \eqref{4.7} and \eqref{4.10}, respectively. Moreover, the quantity $\chi(\al,\beta)$ can be seen as a free parameter of the model to be adjusted to agree with some property of interest of the Boltzmann equation. With the replacement \eqref{5.1}, the BGK-like model for the granular suspension in steady state reads
\beqa
\label{5.2}
& & -aV_y\frac{\partial f}{\partial V_x}-\lambda_t\frac{\partial}{\partial \mathbf{v}}\cdot \mathbf{V} f-\gamma_t \frac{T_{\text{ex}}}{m}\frac{\partial^2 f}{\partial v^2}-\lambda_r\frac{\partial}{\partial \boldsymbol{\omega}}\cdot \boldsymbol{\omega} f\nonumber\\
& & -\gamma_r \frac{T_{\text{ex}}}{I}\frac{\partial^2 f}{\partial \omega^2}=-\chi\nu_t\left(f-f_0\right),
\eeqa
where
\beq
\label{5.3}
\lambda_t\equiv \gamma_t+\frac{\zeta_t}{2}, \quad \lambda_r\equiv \gamma_r+\frac{\zeta_r}{2}.
\eeq

The use of the BGK-like model allows us to determine not only the rheological properties (which are connected with the elements of the pressure tensor) but also all the velocity moments of the velocity distribution function. For a three-dimensional system, it is convenient in the simple shear flow problem to define the general velocity moments
\beq
\label{5.4}
M_{k_1,k_2,k_3}=\int \dd \boldsymbol{\omega} \int \dd \mathbf{V}\;  V_x^{k_1}V_y^{k_2}V_z^{k_3} f(\mathbf{V}, \boldsymbol{\omega}).
\eeq
Note that here we are essentially interested in computing the velocity moments of $f$ involving the translational (peculiar) velocities $\mathbf{V}$. To obtain these moments, we multiply both sides of Eq.\ \eqref{4.1} by $V_x^{k_1}V_y^{k_2}V_z^{k_3}$ and integrate over $\mathbf{V}$ and $\boldsymbol{\omega}$. The result is
\beq
\label{5.5}
a k_1 M_{k_1-1,k_2+1,k_3}+\left(\chi \nu_t+k \lambda_t\right)M_{k_1,k_2,k_3}=N_{k_1,k_2,k_3},
\eeq
where $k=k_1+k_2+k_3$, and
\begin{widetext}
\beq
\label{5.5.1}
N_{k_1,k_2,k_3}=\frac{\gamma_t T_\text{ex}}{m} R_{k_1,k_2,k_3}+\chi \nu_t M_{k_1,k_2,k_3}^\text{L}.
\eeq
The quantities $R_{k_1,k_2,k_3}$ and $M_{k_1,k_2,k_3}^\text{L}$ are defined, respectively, as
\beqa
\label{5.6}
R_{k_1,k_2,k_3}&=&\int \dd \boldsymbol{\omega}\int \dd \mathbf{V}\; f(\mathbf{V}, \boldsymbol{\omega})\frac{\partial^2}{\partial V^2}\left(V_x^{k_1}V_y^{k_2}V_z^{k_3}\right)\nonumber\\
&=&
k_1(k_1-1)M_{k_1-2,k_2,k_3}+k_2(k_2-1)M_{k_1,k_2-2,k_3}+k_3(k_3-1)M_{k_1,k_2,k_3-2},
\eeqa
and
\beq
\label{5.7}
M_{k_1,k_2,k_3}^\text{L}=n \left(\frac{2T_t}{m}\right)^{k/2}\pi^{-3/2}\Gamma\left(\frac{k_1+1}{2}\right)\Gamma\left(\frac{k_2+1}{2}\right)
\Gamma\left(\frac{k_3+1}{2}\right)
\eeq
\end{widetext}
if $k_1$, $k_2$, and $k_3$ are even, being zero otherwise. As expected, the structure of Eq.\ \eqref{5.5} is the same as in the smooth case \cite{GGG19} and hence, the solution to Eq.\ \eqref{5.5} can be written as
\beq
\label{5.9}
M_{k_1,k_2,k_3}=\sum_{q=0}^{k_1} \frac{k_1!}{(k_1-q)!} \frac{(-a)^q}{\left(\chi \nu_t+k \lambda_t\right)^{1+q}}N_{k_1-q,k_2+q,k_3}.
\eeq

The (reduced) nonzero elements of the pressure tensor $P_{k\ell}^*$ and the (reduced) shear rate $a^*$ can be easily obtained from Eqs.\ \eqref{5.5.1}--\eqref{5.9}. Their expressions are
\beq
\label{5.10}
P_{yy}^*=\frac{2+\chi \nu_t^* \sqrt{\theta_t}\theta_t}{2+\left(\chi \nu_t^*+\zeta_t^*\right)\sqrt{\theta_t}},
\eeq
\beq
\label{5.11}
P_{xy}^*=-\frac{2+\chi \nu_t^* \sqrt{\theta_t}\theta_t}{\left[2+\left(\chi \nu_t^*+\zeta_t^*\right)\sqrt{\theta_t}\right]^2}a^*,
\eeq
\beq
\label{5.12}
P_{xx}^*=\frac{2+\chi \nu_t^* \sqrt{\theta_t}\theta_t}{2+\left(\chi \nu_t^*+\zeta_t^*\right)\sqrt{\theta_t}}\Bigg[1+\frac{2a^{*2}}
{\left[2+\left(\chi \nu_t^*+\zeta_t^*\right)\sqrt{\theta_t}\right]^2}\Bigg],
\eeq
\beq
\label{5.13}
a^*=\sqrt{\frac{d}{2}\frac{\sqrt{\theta_t}\zeta_t^*+2(1-\theta_t^{-1})}{\sqrt{\theta_t}\chi \nu_t^*+2\theta_t^{-1}}}\left[2+\sqrt{\theta_t}\left(\chi \nu_t^*+\zeta_t^*\right)\right],
\eeq
where $\zeta_t^*$ is defined by Eq.\ \eqref{4.15} and
\beq
\label{5.14}
\nu_t^*\equiv \frac{\nu_t}{\sqrt{\theta_t}\gamma_t}=\frac{16}{5}\sqrt{\pi}n^*\sqrt{T_\text{ex}^*}.
\eeq
Upon deriving Eqs.\ \eqref{5.10}--\eqref{5.13}, use has been made of the first identity of Eq.\ \eqref{5.3}.

Comparison between Eqs.\ \eqref{4.7}--\eqref{4.19} (derived from Grad's solution to the Boltzmann equation) with Eqs.\ \eqref{5.10}--\eqref{5.13} shows that the BGK results for the non-Newtonian transport properties coincide with the Boltzmann ones when $\chi(\kappa,\al,\beta)$ is chosen as
\beqa
\label{5.15}
\chi=\frac{\nu_\eta^*-\zeta_t^*}{\nu_t^*}&=&\frac{1}{3}\widetilde{\al}\left(1+2\widetilde{\al}\right)+\frac{1}{3}\widetilde{\beta}
\left(1+2\widetilde{\beta}\right) \nonumber\\
& &-2\widetilde{\al}\widetilde{\beta}+\frac{7}{6}\frac{\widetilde{\beta}^2}{\kappa}\frac{\theta_r}{\theta_t}.
\eeqa
We will take this choice for computing the remaining moments of the distribution $f$.

\subsection{Suspension model at $T_\text{ex}=0$ and $\gamma_r=0$}

As happens in the smooth case, \cite{GGG19} in spite of the apparent simplicity of the BGK-like model \eqref{5.2}, it is still intricate to get the explicit form of the velocity distribution function $f(\mathbf{V}, \boldsymbol{\omega})$. In order to obtain $f$ and following the arguments of Ref.\ \onlinecite{S11}, we focus our attention in the \emph{marginal} distribution function
\beq
\label{5.16}
f^{\text{tr}}=\int \dd \boldsymbol{\omega} \; f(\mathbf{V}, \boldsymbol{\omega}).
\eeq
Given that the rheological properties are essentially linked to the translational part of the distribution $f$, one expects that $f^{\text{tr}}$ captures the main properties of the global distribution $f$. Moreover, as in Ref.\ \onlinecite{GGG19}, we also assume the simple limit case $T_\text{ex}=0$, $\gamma_r=0$, but keeping $\gamma_t\equiv \text{const}$. In other words, we are neglecting first the coupling between the rotational degrees of freedom of spheres with the background gas ($\gamma_r=0$). In addition, we are also supposing that $T_\text{ex}$ is much smaller than the translational temperature $T_t$ in such a way that the only relevant effect of the surrounding interstitial gas on grains is through the viscous drag force. In the case of smooth inelastic hard spheres, this simple model has been employed to analyze rheology in sheared granular suspensions, \cite{TK95,SMTK96,SA17,ChVG15,G17,SA20} particle clustering due to hydrodynamic interactions, \cite{WK00} driven steady states, \cite{WZLH09} and to asses the impact of friction in sheared hard-spheres suspensions. \cite{H13,SMMD13,HT13,WGZS14}

The BGK kinetic equation for $f^{\text{tr}}(\mathbf{V})$ can be easily obtained from Eq.\ \eqref{5.2} by integrating over $\boldsymbol{\omega}$: \beq
\label{5.17}
-aV_y\frac{\partial}{\partial V_x}f^{\text{tr}}-\lambda_t\frac{\partial}{\partial \mathbf{v}}\cdot \mathbf{V} f^{\text{tr}}+\chi\nu_t f^{\text{tr}}=\chi\nu_t f_0^{\text{tr}},
\eeq
where
\beq
\label{5.18}
f_0^{\text{tr}}(\mathbf{V})=\int \dd \boldsymbol{\omega} f_0(\mathbf{V},\boldsymbol{\omega})
=n\left(\frac{m}{2\pi T_t}\right)^{3/2} e^{-mV^2/2T_t}.
\eeq
Exploiting the analogy with the smooth case, \cite{GGG19} the hydrodynamic solution to Eq.\ \eqref{5.17} is
\beq
\label{5.19}
f^{\text{tr}}(\mathbf{V})=\int_0^{\infty}\; \dd s\; e^{-(1-3\widetilde{\lambda}_t)s}\;
e^{\widetilde{a}sV_y \frac{\partial}{\partial V_x}}\; e^{\widetilde{\lambda}_ts\mathbf{V}\cdot \frac{\partial}{\partial \mathbf{V}}}
f_0^{\text{tr}}(\mathbf{V}),
\eeq
where $\widetilde{\lambda}_t\equiv \lambda_t/(\chi \nu_t)$ and $\widetilde{a}\equiv a/(\chi \nu_t)$. In Eq.\ \eqref{5.19}, the action of the velocity operators $e^{\widetilde{a}sV_y \frac{\partial}{\partial V_x}}$ and
$e^{\widetilde{\lambda}_t s \mathbf{V}\cdot \frac{\partial}{\partial \mathbf{V}}}$ on an arbitrary function $g(\mathbf{V})$ is
\beq
\label{5.20}
e^{\widetilde{a}sV_y \frac{\partial}{\partial V_x}}g(V_x,V_y,V_z)=g(V_x+\widetilde{a}sV_y, V_y,V_z),
\eeq
\beq
\label{5.25}
e^{\widetilde{\lambda}s\mathbf{V}\cdot \frac{\partial}{\partial \mathbf{V}}}g(V_x,V_y,V_z)=
g\left(e^{\widetilde{\lambda}s}V_x,e^{\widetilde{\lambda}s}V_y,e^{\widetilde{\lambda}s}V_z\right).
\eeq

The elements of the pressure tensor can be computed from the marginal distribution function \eqref{5.19}. They are \cite{G17}
\beq
\label{5.26}
P_{yy}=P_{zz}=\frac{nT_t}{1+2\xi}, \quad P_{xy}=-\frac{nT_t}{(1+2\xi)^2}\widetilde{a},
\eeq
and $P_{xx}=3p-2P_{yy}$. Here, $\xi$ is the real root of the cubic equation $3\xi(1+2\xi)^2=\widetilde{a}^2$. More explicitly, it is given by
\beq
\label{5.27}
\xi(\widetilde{a})=\frac{2}{3}\sinh\Big[\frac{1}{6}\cosh^{-1}\Big(1+\frac{27}{3}\widetilde{a}^2\Big)\Big].
\eeq
The steady balance equation \eqref{3.3} for $T_t$ becomes here $2\gamma_t T_t+\zeta_t T_t=-(2/3n)a P_{xy}$. This equation can be more explicitly written when one takes into account Eqs.\ \eqref{5.26} and \eqref{5.27} with the result
\beq
\label{5.28}
\gamma_t=\chi \nu_t \xi -\frac{1}{2}\zeta_t.
\eeq
Thus, as noted in previous works, \cite{GGG19,G17}  at given values of $\al$, $\beta$, and $\kappa$, the right hand side of Eq.\ \eqref{5.28} vanishes for a certain value $\widetilde{a}_0 (\al, \beta, \kappa)$ of the (reduced) shear rate. Since $\gamma_t$ is strictly positive (except for $a=0$ and $\zeta_t=0$), then physical solutions to \eqref{5.28} are only possible for values of the shear rate $\widetilde{a}$ larger than or equal to $\widetilde{a}_0$. Thus, in particular, when $\al \neq 1$ or $|\beta| \neq 1|$, the constraint \eqref{5.28} prevents the possibility of obtaining the Navier--Stokes shear viscosity (i.e., when $\widetilde{a}\to 0$) of the granular suspension. This is in fact a drawback of this simple model not shared by the generalized Fokker--Planck suspension model introduced in Sec.\ \ref{sec2}. In the case of smooth inelastic hard spheres, this drag model has been widely used for many authors \cite{TK95,SA17,SA20} to study the discontinuous transition for the temperature between the \emph{quenched} and the \emph{ignited} states.

\section{Rheology of sheared dry granular gases}
\label{sec6}

Although the main goal of this paper is to assess the influence of the interstitial gas on the rheological properties of inelastic rough hard spheres, it is interesting first to analyze the results obtained in the dry limit case (namely, when the effect of the background gas is neglected). To the best of our knowledge, this problem was independently studied many years ago for moderately dense gases by Jenkins and Richman\cite{JR85a} for hard disks and by Lun \cite{L91} for hard spheres. In both works, the calculations were in principle restricted to nearly elastic collisions ($\al \lesssim 1)$ and either nearly smooth particles ($\beta \lesssim 1$) or nearly perfectly rough spheres ($\beta \gtrsim -1$). A more recent study has been performed by Santos \cite{S11} by using the BGK-like kinetic model defined in Eq.\ \eqref{5.1}. Given that the BGK results for rheology agree with those derived by solving the Boltzmann equation from Grad's moment method, only a comparison with the theoretical predictions reported by Lun \cite{L91} for a three-dimensional gas will be offered in this Section.

\begin{figure}
\includegraphics[width=0.8\columnwidth]{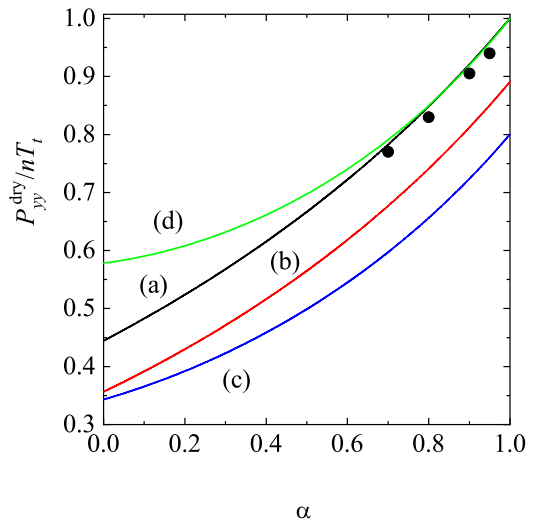}
\caption{Plot of the (reduced) element $P_{yy}^\text{dry}/nT_t$ as a function of the coefficient of normal restitution $\al$ for $\kappa=\frac{2}{5}$ and four different values of the coefficient of tangential restitution $\beta$: $\beta=-1$ (a), $\beta=-0.5$ (b), $\beta=0.5$ (c), and $\beta=1$ (d). Symbols refer to DSMC results obtained for spheres perfectly smooth ($\beta=-1$).\cite{MG02a} Reproduced with permission from Physica A \textbf{310}, 17 (2002). Copyright 2002 Elsevier.}
\label{fig3}
\end{figure}
\begin{figure}
\includegraphics[width=0.8\columnwidth]{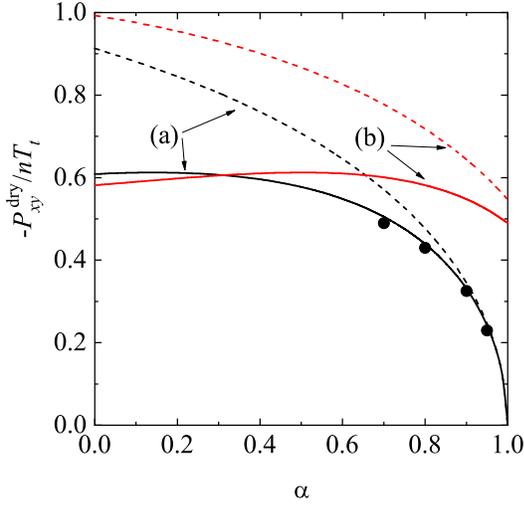}
\caption{Plot of the (reduced) element $-P_{xy}^\text{dry}/nT_t$ as a function of the coefficient of normal restitution $\al$ for $\kappa=\frac{2}{5}$ and two different values of the coefficient of tangential restitution $\beta$: $\beta=-1$ (a) and $\beta=0.5$ (b). The solid lines correspond to the results obtained here while the dashed lines refer to the results derived by Lun.\cite{L91} Symbols refer to DSMC results obtained for spheres perfectly smooth ($\beta=-1$).\cite{MG02a} Reproduced with permission from Physica A \textbf{310}, 17 (2002). Copyright 2002 Elsevier.}
\label{fig4}
\end{figure}

A way of obtaining the results for the dry case consists in formally setting $\gamma_t=\gamma_r=0$. However, one has to take care in extracting the results for the dry case from those derived in Sec.\ \ref{sec5} since practically all of them have been expressed in terms of dimensionless quantities that diverge when $\gamma_t\to 0$. Thus, one has to solve first the set \eqref{4.11} for the nonzero elements $P_{yy}^\text{dry}$ and $P_{xy}^\text{dry}$ (recall that $P_{xx}^\text{dry}=3p-2P_{yy}^\text{dry}$) and then substitute these forms into the balance equation \eqref{3.3}. After some simple algebra, one simply gets
\beq
\label{6.1}
P_{yy}^\text{dry}=P_{zz}^\text{dry}=1-\frac{\zeta_t}{\nu_\eta}, \quad P_{xy}^\text{dry}=-\frac{P_{yy}^\text{dry}}{\nu_\eta}a,
\eeq
\beq
\label{6.2}
a^2=\frac{3}{2}\frac{\zeta_t \nu_\eta}{P_{yy}^\text{dry}},
\eeq
where $\nu_\eta$ and $\zeta_t$ are given by Eqs.\ \eqref{4.6} and \eqref{4.7}, respectively. Finally, the ratio of the the rotational to translational temperature can be easily obtained from the balance equation \eqref{3.4} by taking $\gamma_r=0$. It leads to the condition $\zeta_r=0$, which according to Eq.\ \eqref{4.10} yields
\beq
\label{6.3}
\Big(\frac{T_r}{T_t}\Big)^{\text{dry}}=\kappa \frac{1+\beta}{1-\beta+2\kappa}.
\eeq
Equation \eqref{6.3} was already obtained by Lun.\cite{L91} As happens in the homogenous steady state driven by a white-noise thermostat,\cite{VS15} the temperature ratio of the steady shear flow problem is independent of the coefficient of restitution $\al$. This conclusion contrasts with the results derived in the homogeneous cooling case, \cite{HZ97,LHMZ98,HHZ00,Z06} where $T_r/T_t$ depends on both $\al$ and $\beta$ [see Eq.\ \eqref{b3} of the Appendix \ref{appB}.]

Contrary to the case of granular suspensions, the balance equation \eqref{3.3} establishes an intrinsic relation between the (reduced) shear rate $a/\nu_t$ and the mechanical parameters of the system (the coefficients of normal restitution $\al$ and tangential restitution $\beta$ and the dimensionless moment of inertia $\kappa$). This means that $a/\nu_t$ is not an independent parameter and is a function of $\al$, $\beta$, and $\kappa$.

\begin{figure}[h!]
\centering
\includegraphics[width=0.8\columnwidth]{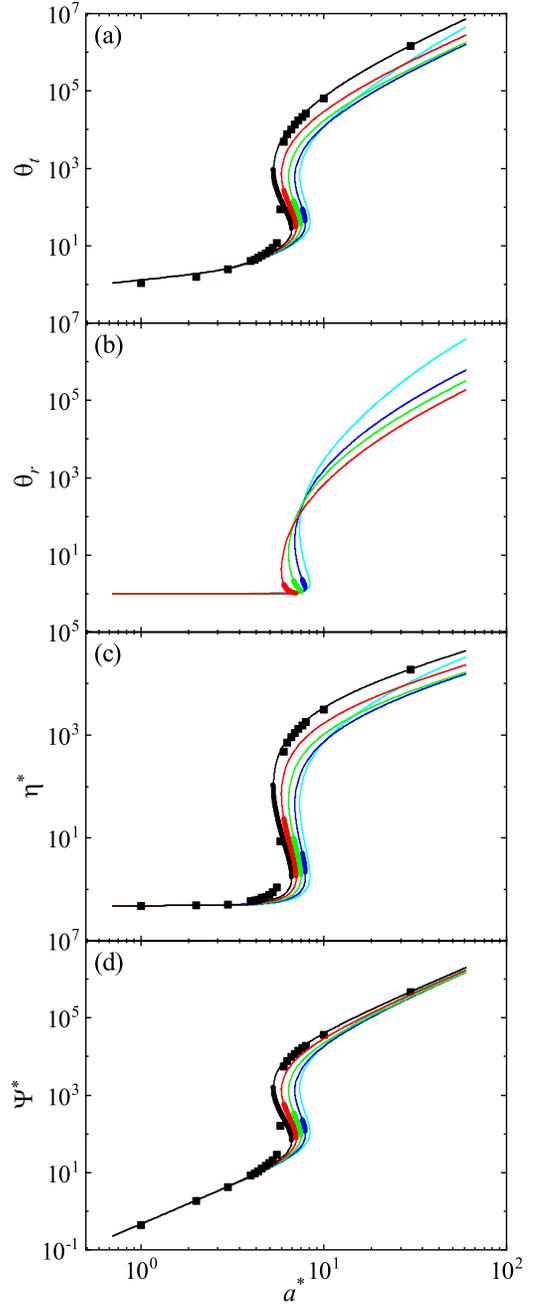}
\caption{Plots of the (steady) translational granular temperature $\theta_t$ (panel (a)), the (steady) rotational granular temperature $\theta_r$ (panel (b)), the non-Newtonian shear viscosity $\eta^*$ (panel (c)), and the viscometric function $\Psi^*$ (panel (d)) as a function of the (reduced) shear rate $a^*$ for $\al=0.9$ and different values of the coefficient of tangential restitution $\beta$: $\beta=-1$ (black line), $\beta=-0.5$ (red line), $\beta=0$ (green line), $\beta=0.5$ (blue line), and $\beta=1$ (cyan line). Here, $\kappa=\frac{2}{5}$, $n^*=0.01$, and $T_\text{ex}^*=1$. The thick lines represent the linearly unstable regions.
Symbols refer to computer simulation results obtained for spheres perfectly smooth ($\beta=-1$).\cite{HTG17}
Reproduced with permission from Phys. Rev. E \textbf{96}, 042903 (2017). Copyright 2017 American Physical Society.
}
\label{fig5}
\end{figure}

Since the results derived by Lun\cite{L91} apply in principle to slightly inelastic, slightly rough spheres, then the normal stress differences vanish: $P_{xx}^\text{dry}=P_{yy}^\text{dry}=P_{zz}^\text{dry}=p$. On the other hand, his expressions for $P_{xy}^\text{dry}$ and $a/\nu_t$ are formally equivalent to our results when one takes $P_{yy}^\text{dry}=1$ in Eqs.\ \eqref{6.1} and \eqref{6.2}. Figure \ref{fig3} shows the $\al$-dependence of the (reduced) $yy$-element $P_{yy}^\text{dry}/nT_t$ for $\kappa=\frac{2}{5}$ and four different values of $\beta$: $\beta=-1$ (perfectly smooth spheres), $\beta=-0.5$ (moderate roughness), $\beta=0.5$ (medium roughness), and $\beta=1$ (strong roughness). Results obtained from DSMC simulations\cite{MG02a} for perfectly smooth spheres are also included. It is quite apparent first that the combined effect of $\al$ and $\beta$ gives rise to anisotropic effects in the $yy$-element of the pressure tensor; these effects are measured by the departure of the ratio $P_{yy}^\text{dry}/nT_t$ from 1. We also see that, for a given value of  $\beta$, these non-Newtonian effects increase monotonically with decreasing $\al$. In addition, for a given value of $\al$, $P_{yy}^\text{dry}/nT_t$ presents a non-monotonic dependence on $\beta$; the impact of roughness being higher for central values of $\beta$ (let's say $|\beta|\backsim 0.5$). Comparison with Monte Carlo simulations for $\beta=-1$ shows a good agreement; we hope that this agreement is also extended for the remaining values of $\beta$. As a complement of Fig.\ \ref{fig3}, Fig. \ref{fig4} plots $-P_{xy}^\text{dry}/nT_t$ versus $\al$ for $\beta=-1$ and $\beta=0.5$. The theoretical predictions of Lun\cite{L91} are also represented. As expected, we observe that the agreement between Lun's predictions and our results is excellent for $\al \lesssim 1$ and $|\beta| \lesssim 1$. On the other hand, the discrepancies between both theories increase as increasing inelasticity (at a given value of roughness) or as increasing roughness (at a given value of inelasticity). As in the case of Fig.\ \ref{fig3}, Fig.\ \ref{fig4} highlights again the good performance of Grad's solution when $\al=1$ and $\beta=-1$ since the above solution compares very well with simulations.

\section{Rheology and fourth-degree moments of sheared inertial suspensions}
\label{sec7}

We consider now sheared inertial suspensions ($\gamma_t\neq 0$ and $\gamma_r \neq 0$). In Sec.\ \ref{sec4} we have determined the elements of the (reduced) pressure tensor $P_{k\ell}^*$ by solving the Boltzmann equation \eqref{3.2} by means of Grad's moment method. Then, in Sec.\ \ref{sec5} we have replaced the Boltzmann collision operator $J[f,f]$ by the BGK-like collision term \eqref{5.1} and have explicitly obtained all the velocity moments of the velocity distribution function. In dimensionless form, all the above quantities (pressure tensor and higher degree velocity moments) have been expressed in terms of the restitution coefficients $\al$ and $\beta$, the (reduced) moment of inertia $\kappa$, the reduced density $n^*$, the (reduced) bath temperature $T_\text{ex}^*$, and the (reduced) shear rate $a^*$.

We want essentially assess the shear-rate dependence of $\eta^*$, $\Psi^*$, $\theta_t$, $\theta_r$, and the fourth-degree moments for fixed values of $\al$, $\beta$, $\kappa$, $n^*$, and $T_\text{ex}$. Since the theoretical results for $\theta_t$, $\eta^*$, and $\Psi^*$ will be compared against event-driven simulations \cite{HTG17} carried out for the case $\al=0.9$ and $\beta=-1$, the values of $n^*$ and $T_\text{ex}$ employed in those simulations ($n^*=0.01$ and $T_\text{ex}=1$) and the value $\kappa=\frac{2}{5}$ will be used in the remaining plots of this Section.

\subsection{Rheology}

The dependence of the (steady) translational temperature $\theta_t$, the non-Newtonian shear viscosity $\eta^*$, and the viscometric function $\Psi^*$ on the (reduced) shear rate $a^*$ is shown in Fig.\ \ref{fig5}. The analytical forms of the above quantities are given by Eqs.\ \eqref{4.19}, \eqref{4.21}, and \eqref{4.22}, respectively. We recall that the corresponding expressions of the BGK equation agree with those derived from Grad's solution when one makes the choice \eqref{5.15} for the free parameter $\chi$ of the kinetic model. In addition, as will be discussed in Sec.\ \ref{sec8}, depending on the values of $\al$ and $\beta$, the steady solution can be linearly unstable. The thick lines in Fig.\ \ref{fig5} denote the linearly unstable regions.

The main conclusion of Fig.\ \ref{fig5} is that the roughness does not change the trends observed in previous works\cite{HTG17,HTG20,GGG19} for perfectly smooth inelastic spheres: there is a drastic increase of all the rheological properties with increasing the shear rate. In particular, the panel (c) of Fig.\ \ref{fig5} highlights the existence of DST for the shear viscosity $\eta^*$, regardless of the value of the coefficient of restitution $\beta$. On the other hand, at a more quantitative level, we observe that, for a given value of $a^*$, high levels of roughness can slightly attenuate the jump of $\eta^*$ relative to the frictionless case. This is a quite unexpected result since most of the results obtained for concentrated suspensions have shown that friction enhances DST. However, this trend is not monotonic since there is a change in the above behavior for very high shear rates; in fact, the line corresponding to strong roughness ($\beta=1$) intersects the curves of $\beta=0.5$, $\beta=0$, and $\beta=-0.5$ for $a^* \gtrsim 10$. In addition, the agreement between theory and simulations for perfectly smooth spheres ($\al=0.9$ and $\beta=-1$) is relatively good, except in a small region close to the transition point where simulation data suggest a sharper transition than the Boltzmann one. We think that this small discrepancy is mainly due to the limitations of the Boltzmann equation for accounting small density corrections to $\eta^*$ around this transition point. As a matter of fact, the Enskog predictions for this quite small density ($n^*=0.01$) compares slightly better with simulation data than the ones obtained from the Boltzmann equation; see for instance Fig.\ 2 of Ref.\ \onlinecite{HTG17}.

Although similar trends are observed for $\theta_t$ and $\Psi^*$, it is worthwhile noticing that the combined effect of $\al$ and $\beta$ on the viscometric function $\Psi^*$ is quite important since while this quantity is tiny for small shear rates, it suddenly increases for not quite large values of the shear rate (let's say $a^*\thickapprox 1$). It must be recalled that the results obtained in the context of the Enskog equation for moderately dense gases have shown a transition from DST for very dilute suspensions to CST at relatively moderate densities.\cite{HTG17,HTG20}

More influence of roughness on rheology can be found in the case of the (steady) rotational granular temperature $\theta_r=T_r/T_\text{ex}$. This quantity does not play any role in the perfectly smooth case. The panel (b) of Fig.\ \ref{fig5} shows the shear-rate dependence of $\theta_r$. It is quite apparent that, for large shear rates, roughness clearly enhances the value of $\theta_r$ in contrast to what happens for $\theta_t$. It must be remarked that similar features of the rheological properties have been observed for other values of the coefficient of restitution.

\subsection{Fourth-degree velocity moments}

We consider now the relevant fourth-degree velocity moments obtained in the context of the BGK model. They can be easily determined from Eq.\ \eqref{5.9}. As discussed in Ref.\ \onlinecite{GGG19}, there are eight independent fourth-degree (symmetric) moments: five of them are even functions of the (reduced) shear rate $a^*$ while the remaining three are odd functions of $a^*$. To illustrate the shear-rate dependence of those moments, we chose the representative moments
\beqa
\label{7.1}
M_{4|0}&=&\int \dd{\bf v}\int \dd\boldsymbol{\omega} \; V^4\; f(\boldsymbol{\omega},\mathbf{V})\nonumber\\
&=&M_{400}+2\left(M_{040}+M_{220}+M_{202}+M_{022}\right),\nonumber\\
\eeqa
\beqa
\label{7.2}
M_{2|xy}&=&\int \dd{\bf v}\int \dd\boldsymbol{\omega} \; V^2 V_x V_y f(\boldsymbol{\omega},\mathbf{V})\nonumber\\
&=&M_{310}+M_{130}+M_{112},
\eeqa
\begin{figure}[h!]
\centering
\includegraphics[width=0.8\columnwidth]{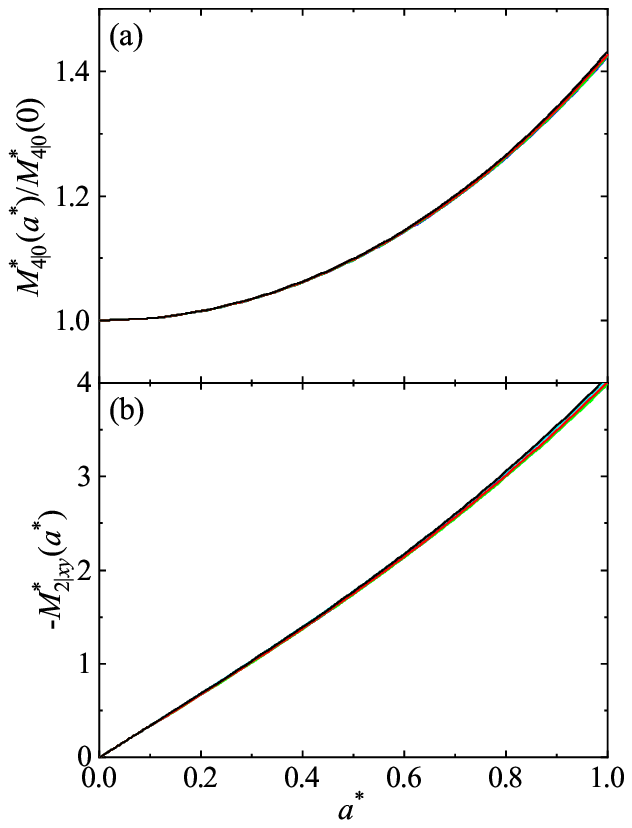}
\caption{Shear-rate dependence of the (scaled) fourth-degree moments $M_{4|0}^*(a^*)/M_{4|0}^*(0)$ (panel (a)) and $-M_{2|xy}^*(a^*)$ (panel (b)) for $\al=0.9$ and different values of the coefficient of tangential restitution $\beta$: $\beta=-1$ (black line), $\beta=-0.5$ (red line), $\beta=0$ (green line), $\beta=0.5$ (blue line), and $\beta=1$ (cyan line). Here, $\kappa=\frac{2}{5}$, $n^*=0.01$, and $T_\text{ex}^*=1$.}
\label{fig6}
\end{figure}
\begin{figure}[h!]
\centering
\includegraphics[width=0.8\columnwidth]{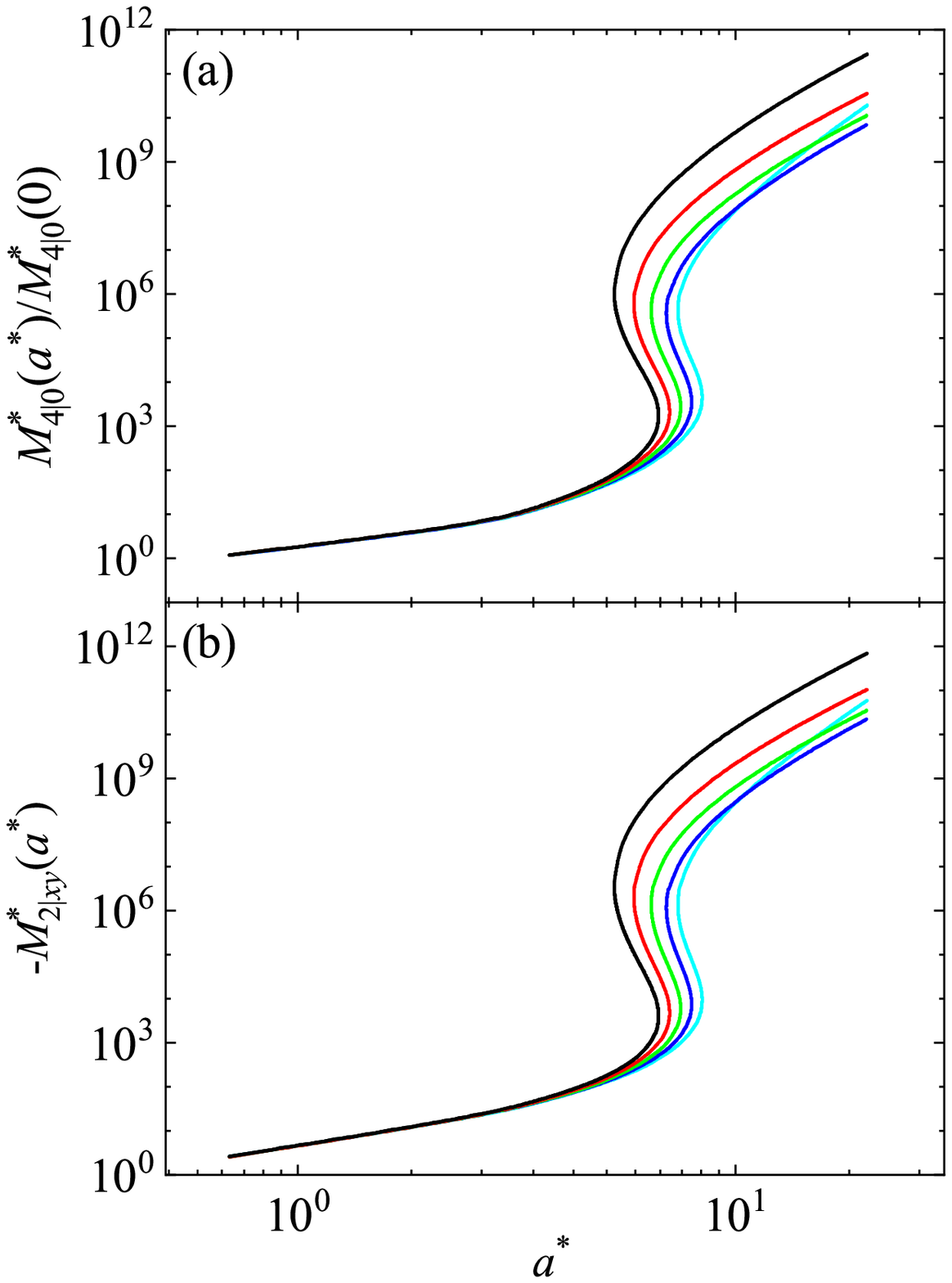}
\caption{Shear-rate dependence of the (scaled) fourth-degree moments $M_{4|0}^*(a^*)/M_{4|0}^*(0)$ (panel (a)) and $-M_{2|xy}^*(a^*)$ (panel (b)) for $\al=0.9$ and different values of the coefficient of tangential restitution $\beta$: $\beta=-1$ (black line), $\beta=-0.5$ (red line), $\beta=0$ (green line), $\beta=0.5$ (blue line), and $\beta=1$ (cyan line). Here, $\kappa=\frac{2}{5}$, $n^*=0.01$, and $T_\text{ex}^*=1$.}
\label{fig7}
\end{figure}
where the canonical moments $M_{k_1,k_2,k_3}$ are given by Eq.\ \eqref{5.9}. Upon writing Eq.\ \eqref{6.1}, use has been made of the identity $M_{040}=M_{004}$. While the moment $M_{4|0}$ is an even function of $a^*$ (and so, $M_{4|0}\neq 0$ when $a^*=0$), the moment $M_{2|xy}$ is an odd function of $a^*$ (and so, $M_{2|xy}=0$ when $a^*=0$). To see more clearly the influence of both $\al$ and $\beta$ on $M_{4|0}$ and $M_{2|xy}$, we consider first the region $0\leqslant a^* \leqslant 1$ where non-Newtonian effects are expected to be still important. Figure \ref{fig6} shows the shear-rate dependence of $M_{4|0}^*(a^*)/M_{4|0}^*(0)$ and $-M_{2|xy}^*(a^*)$ for $\al=0.9$ and several values of $\beta$. Here, we have introduced the dimensionless moments
\beq
\label{7.3}
\left\{M_{4|0}^*, M_{2|xy}^*\right\}=n^{-1} \left(\frac{m}{T_\text{ex}}\right)^2\left\{M_{4|0}, M_{2|xy}\right\}.
\eeq
In Fig.\ \ref{fig6}, $M_{4|0}^*(0)$ refers to the value of $M_{4|0}^*$ when $a^*=0$, namely,
\beqa
\label{7.4}
M_{4|0}^*(0)&=&\frac{9}{4+\sqrt{\theta_t^{(0)}}\left(\chi \nu_t^*+2\zeta_t^*\right)}\Big[
\chi \nu_t^*\sqrt{\theta_t^{(0)}}\theta_t^{(0)}
\nonumber\\
& & +4\frac{2+\chi \nu_t^*\sqrt{\theta_t^{(0)}}\theta_t^{(0)}}{2+\sqrt{\theta_t^{(0)}}\left(\chi \nu_t^*+\zeta_t^*\right)}\Big],
\eeqa
where $\theta_t^{(0)}$ is a real solution of Eq.\ \eqref{4.23}. As expected, we observe first in Fig.\ \ref{fig6} that these fourth-degree moments clearly depart from their equilibrium values (in the absence of shear rate). Surprisingly, at a given value of $\al$, the impact of $\beta$ on those moments is very small since all the curves collapse in a common one. This feature contrasts with the results obtained for the rheological properties since the effect of $\beta$ on both $\eta^*$ and $\Psi^*$ is remarkable in this range of values of the shear rate ($a^*\leq 1$). It must be recalled that a similar property appears in the smooth limit case, \cite{GGG19} since the effect of $\al$ on $M_{4|0}^*(a^*)$ and $-M_{2|xy}^*(a^*)$ was also found very tiny at a given value of the shear rate.

For very large values of the shear rate, it is interesting to see whether the fourth-degree moments increase also dramatically with the shear rate in a similar way as the non-Newtonian shear viscosity $\eta^*$. This is illustrated in Fig.\ \ref{fig7} where it is clearly shown that both scaled moments exhibit an $S$-shape for any value of $\beta$. In addition, we also see that the effect of $\beta$ on these moments is really significant for large values of $a^*$.

\begin{figure}[h!]
\includegraphics[width=0.8\columnwidth]{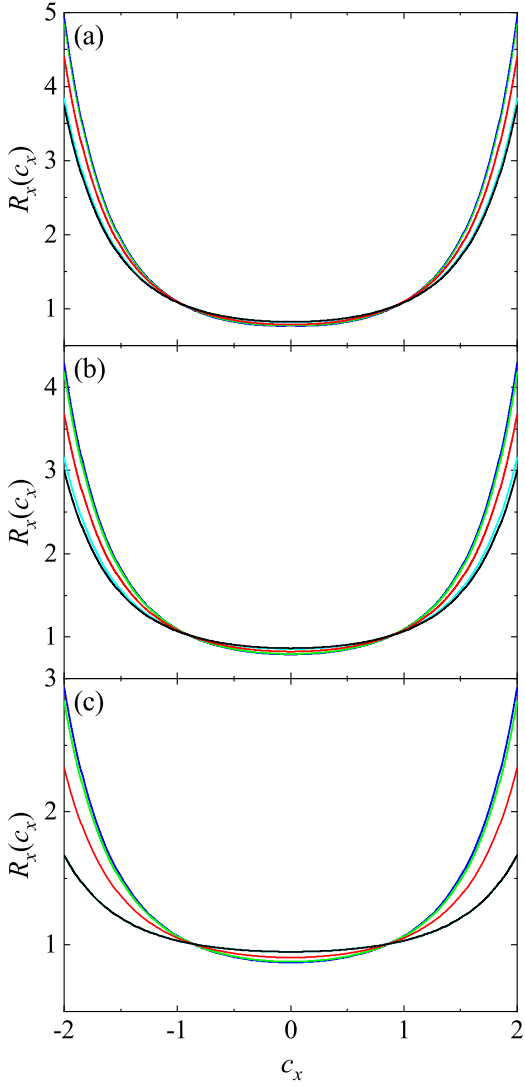}
\caption{Plot of the ratio $R_x(c_x)=\varphi(c_x)/(\pi^{-1/2}e^{-c_x^2})$ versus the scaled velocity $c_x=\sqrt{m/2T_t}V_x$  for
$\widetilde{\gamma}_t=0.1$ and five different values of the coefficient of tangential restitution $\beta$: $\beta=-1$ (black line), $\beta=-0.5$ (red line), $\beta=0$ (green line), $\beta=0.5$ (blue line), and $\beta=1$ (cyan line). Three different values of the coefficient of normal restitution $\al$ are considered: $\al=1$ (panel (a)), $\al=0.7$ (panel (b)), and $\al=0.5$ (panel (c)). Here, $\kappa=\frac{2}{5}$, $n^*=0.01$, and $T_\text{ex}^*=1$.}
\label{fig7.1}
\end{figure}

\subsection{Velocity distribution function}

As said in Sec.\ \ref{sec5}, one of the main practical advantages of kinetic models is the possibility of obtaining the explicit form of the velocity distribution function. Here, we have obtained it in the special case $T_\text{ex}=\gamma_r=0$ and is given by
Eq.\ \eqref{5.19}. To illustrate the dependence of $f^{\text{tr}}(\mathbf{V})$ on the parameter space of the problem, let us rewrite this distribution as
\beq
\label{7.5}
f^{\text{tr}}(\mathbf{V})=n\Big(\frac{m}{2T_t}\Big)^{3/2}\varphi(\mathbf{c}),
\eeq
where $\mathbf{c}=\sqrt{m/2T_t}\mathbf{V}$ is the reduced peculiar velocity and the reduced velocity distributions function $\varphi(\mathbf{c})$ is given by
\beqa
\label{7.6}
\varphi(\mathbf{c})&=&\pi^{-d/2}\int_0^{\infty}\; \dd s\;  e^{-(1-3\widetilde{\lambda}_t)s}\exp\Big\{-e^{2\widetilde{\lambda}_t s}\;
\nonumber\\
& & \times
\big[(c_x+\widetilde{a}s c_y)^2+c_y^2+c_z^2)\big]\Big\}.
\eeqa
Upon writing Eq.\ \eqref{7.6}, use has been made of Eqs.\ \eqref{5.20} and \eqref{5.25}. Figure \ref{fig7.1} shows the ratio $R_x(c_x)=\varphi_x(c_x)/\big(\pi^{-1/2}e^{-c_x^2}\big)$ for $\widetilde{\gamma}_t=0.1$ and different values of the restitution coefficients $\al$ and $\beta$. Here, $\varphi_x(c_x)$ is the marginal distribution
\beqa
\label{7.7}
\varphi_x(c_x)&=&\int_{-\infty}^{\infty}\; \dd c_y\;\int_{-\infty}^{\infty}\; \dd c_z\; \varphi(\mathbf{c})\nonumber\\
&=&\frac{1}{\sqrt{\pi}}\int_0^{\infty}\; \dd s \frac{e^{-(1-\widetilde{\lambda}_t)s}}{\sqrt{1+\widetilde{a}^2s^2}}\text{exp}
\left(-e^{2\widetilde{\lambda}_ts} \frac{c_x^2}{1+\widetilde{a}^2s^2}\right).\nonumber\\
\eeqa
Figure \ref{fig7.1} shows that in general $R_x(c_x)$ is clearly different from 1, namely, the distribution $\varphi_x(c_x)$ is highly distorted from its local equilibrium value ($\pi^{-1/2}e^{-c_x^2}$). At a given value of the coefficient of tangential restitution $\beta$, the distortion is more significant as the coefficient of normal restitution $\al$ decreases (increasing inelasticity). The impact of roughness on $R_x(c_x)$ increases with decreasing $\al$.

\section{Linear stability analysis of the steady solution}
\label{sec8}

Although our study has been mainly focused on the determination the rheological properties under steady state conditions, an interesting question is to see if actually the steady state solution provided by Eqs.\ \eqref{4.17}--\eqref{4.20} is indeed a (linearly) stable solution. In order to perform this analysis, we write first the four relevant equations for $P_{yy}^*$, $P_{xy}^*$, $\theta_t$, and $\theta_r$ from Eq.\ \eqref{4.16.1}:
\beq
\label{a1}
\partial_\tau P_{yy}^*+2\left(P_{yy}^*-1\right)=-\nu_\eta^*\sqrt{\theta_t}\left(P_{yy}^*-\theta_t\right)-\sqrt{\theta_t}\theta_t \zeta_t^*,
\eeq
\beq
\label{a2}
\partial_\tau P_{xy}^*+a^* P_{yy}^*+2 P_{xy}^*=-\nu_\eta^*\sqrt{\theta_t}P_{xy}^*,
\eeq
\beq
\label{a3}
\partial_\tau \theta_t+2\left(\theta_t-1\right)+\sqrt{\theta_t}\theta_t \zeta_t^*=-\frac{2}{3}a^* P_{xy}^*,
\eeq
\beq
\label{a4}
\partial_\tau \theta_r+2\frac{\gamma_r}{\gamma_t}\left(\theta_r-1\right)+\sqrt{\theta_t}\theta_r \zeta_r^*=0.
\eeq
We want to solve the set of Eqs.\ \eqref{a1}--\eqref{a5} by assuming small deviations from the steady state solution. Thus, we write
\beq
\label{a5}
P_{yy}^*(\tau)=P_{yy,s}^*+\delta P_{yy}^*(\tau), \quad P_{xy}^*(\tau)=P_{xy,s}^*+\delta P_{xy}^*(\tau),
\eeq
\beq
\label{a6}
\theta_{t}(\tau)=\theta_{t,s}+\delta \theta_{t}(\tau), \quad \theta_{r}(\tau)=\theta_{r,s}+\delta \theta_{r}(\tau),
\eeq
where the subscript $s$ means that the quantity is evaluated in the steady state. Here, for the sake of simplicity, we have assumed that the interstitial fluid is not perturbed and hence, the parameters $\gamma_t$, $\gamma_r$, and $T_\text{ex}$ are constant in the time-dependent shear flow problem. This means that the reduced shear rate $a^*=a/\gamma_t$ is also constant. Substituting Eqs.\ \eqref{a5} and \eqref{a6} into Eqs.\ \eqref{a1}--\eqref{a5} and neglecting nonlinear terms in the perturbations, after some algebra one gets the set of linear differential equations
\begin{widetext}
\beq
\label{a7}
\partial_\tau
\left(
\begin{array}{c}
\widetilde{P}_{yy}\\
\widetilde{P}_{xy}\\
\widetilde{\theta}_{t}\\
\widetilde{\theta}_{r}
\end{array}
\right)=-\mathsf{L}\cdot
\left(
\begin{array}{c}
\widetilde{P}_{yy}\\
\widetilde{P}_{xy}\\
\widetilde{\theta}_{t}\\
\widetilde{\theta}_{r}
\end{array}
\right),
\eeq
where
\beq
\label{a8}
\widetilde{P}_{yy}(\tau)=\frac{\delta P_{yy}^*(\tau)}{P_{yy,s}^*}, \quad \widetilde{P}_{xy}(\tau)=\frac{\delta P_{xy}^*(\tau)}{P_{xy,s}^*}, \quad \widetilde{\theta}_{t}(\tau)=\frac{\delta \theta_{t}(\tau)}{\theta_{t,s}}, \quad \widetilde{\theta}_{r}(\tau)=\frac{\delta \theta_{r}(\tau)}{\theta_{r,s}}.
\eeq
The square matrix $\mathsf{L}$ is
\beq
\label{a10}
\mathsf{L}=
\left(
\begin{array}{cccc}
2+\sqrt{\theta_t}\nu_\eta^*&0&\sqrt{\theta_t}\theta_{t}\left(\frac{\nu_\eta^*-2\bar{\nu}_\eta}{2\theta_t}-\frac{
\frac{3}{2}\nu_\eta^*-\bar{\nu}_\eta}{P_{yy}^*}+\frac{\frac{3}{2}\zeta_t^*-\bar{\zeta}_t}{P_{yy}^*}\right)&\sqrt{\theta_t}
\left(\bar{\nu_\eta}-\frac{11 \bar{\nu}_\eta}{P_{yy}^*}\theta_t\right)\\
a^*\frac{P_{yy}^*}{P_{xy}^*}&2+\sqrt{\theta_t}\nu_\eta^*&\sqrt{\theta_t}\left(\frac{1}{2}\nu_\eta^*-\bar{\nu}_\eta\right)&
\sqrt{\theta_t}\bar{\nu}_\eta\\
0&\frac{2}{3}\frac{P_{xy}^*a^*}{\theta_t}&\sqrt{\theta_t}\left(\frac{3}{2}\zeta_t^*-\bar{\zeta}_t\right)+2&\sqrt{\theta_t}\bar{\zeta}_t\\
0&0&\sqrt{\theta_t}\left(\frac{1}{2}\zeta_r^*+\bar{\zeta}_r\right)&2\frac{\gamma_r}{\gamma_t}+
\sqrt{\theta_t}\left(\zeta_r^*-\bar{\zeta}_r\right)
\end{array}
\right),
\eeq
\end{widetext}
where the subscript $s$ has been omitted for the sake of brevity. This means that it is understood that all the quantities appearing in the matrix $\mathsf{L}$ are evaluated at the steady state. In Eq.\ \eqref{a10}, we have introduced the quantities
\beq
\label{a14}
\bar{\nu}_\eta=\frac{8}{15}\sqrt{\pi}\frac{\widetilde{\beta}^2}{\kappa}\frac{\theta_r}{\theta_t}n^*\sqrt{T_\text{ex}^*}, \quad
\bar{\zeta}_t=-10\bar{\nu}_\eta, \quad \bar{\zeta}_r=\frac{\theta_t}{\theta_r}\bar{\zeta}_t.
\eeq
In the purely smooth case ($\beta=-1$), $\widetilde{\beta}=\bar{\nu}_\eta=\bar{\zeta}_t=\bar{\zeta}_r=0$, and hence the matrix $\mathsf{L}$ is consistent with the one obtained in Ref.\ \onlinecite{HT19} for a linear stability analysis for smooth hard spheres.\footnote{The only difference between both results is the existence of some contributions proportional to $(\partial a^*/\partial \theta_t)_s$ in Ref.\ \onlinecite{HT19} which are absent in our results. However, since $a^*\propto T_\text{ex}^{-1/2}\equiv \text{const.}$ in the time-dependent problem, these type of contributions are not present when one slightly perturbs the steady state by small homogeneous time-dependent perturbations.}

\begin{figure}[h!]
\includegraphics[width=0.8\columnwidth]{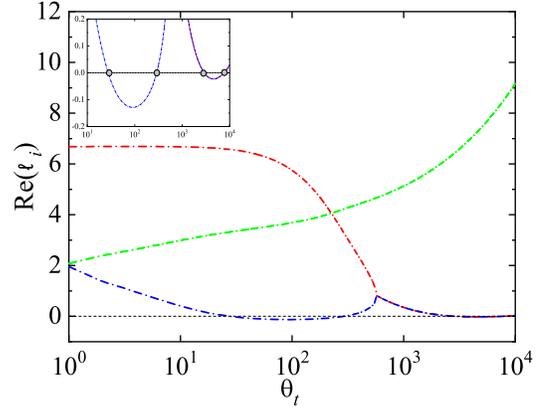}
\caption{Plot of the real part of the eigenvalues $\ell_i$ $(i=1,2,3,4)$ of the matrix $\mathsf{L}$ for $n^*=0.01$, $T_\text{ex}^*=1$,
$\kappa=\frac{2}{5}$, $\al=1$, and $\beta=-0.5$. The green line corresponds to the real part of the complex conjugate pair ($\ell_2,\ell_3$). The red and blue lines refer to the other two eigenvalues ($\ell_1,\ell_4$), which become a complex conjugate pair for high values of $\theta_t$. The region where the real parts of $\ell_1$ and $\ell_4$ vanish is shown more clearly in the inset graph.}
\label{fig8.1}
\end{figure}

\begin{figure}[h!]
\includegraphics[width=0.8\columnwidth]{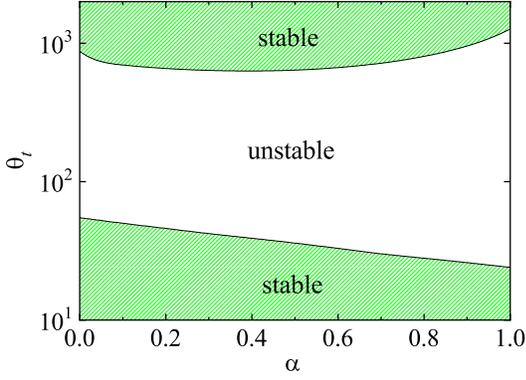}
\caption{Phase diagram for the behavior of the eigenvalues of the matrix $\mathsf{L}$ in the case of purely smooth granular gases ($\beta=-1$) for $n^*=0.01$, $T_\text{ex}^*=1$, and $\kappa=\frac{2}{5}$. The hatched regions correspond to states where the steady simple shear flow solution is linearly stable, while the unfilled region refers to states where the steady solution is linearly unstable.}
\label{fig8.2}
\end{figure}
\begin{figure}[h!]
\includegraphics[width=0.8\columnwidth]{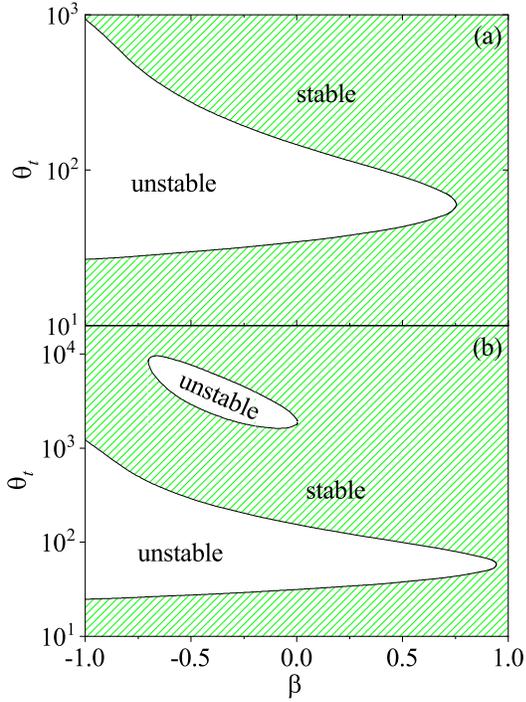}
\caption{Phase diagram for the behavior of the eigenvalues of the matrix $\mathsf{L}$ for $n^*=0.01$, $T_\text{ex}^*=1$, $\kappa=\frac{2}{5}$, and two different values of $\al$: $\alpha=0.9$ (panel (a)) and $\alpha=1$ (panel (b)). The hatched regions correspond to states where the steady simple shear flow solution is linearly stable, while the unfilled regions refer to states where the steady solution is linearly unstable.}
\label{fig8.3}
\end{figure}

The eigenvalues $\ell$ of the square matrix $\mathsf{L}$ govern the time evolution of the deviations $\left\{\widetilde{P}_{yy}, \widetilde{P}_{xy}, \widetilde{\theta}_{t}, \widetilde{\theta}_{r} \right\}$ from the steady solution given by the set $\left\{P_{yy,s}^*,P_{xy}^*, \theta_{t,s}, \theta_{r,s}\right\}$. If the real parts of those eigenvalues are
positive the steady solution is linearly stable, while it is unstable otherwise.

On the other hand, as already occurs for smooth spheres, \cite{GGG19} the (steady) translational temperature $\theta_t(a^*)$ turns out to be a multi-valued function of the (reduced) shear rate in a certain interval of values of $a^*$ (see the vicinity of the saddle point in Fig.\ \ref{fig5}). Thus, as already did in Refs.\ \onlinecite{HT19,GGG19}, in order to analyze the stability of the steady solution we take $\theta_t$ as independent parameter instead of $a^*$ for the sake of convenience. Of course, once $\theta_t(a^*)$ is known, $a^*$ can be determined from Eq.\ \eqref{4.19}. As expected from the previous stability analysis performed for smooth spheres,\cite{HT19} a careful analysis of the eigenvalues $\ell$ shows that, for given values of $\al$ and $\beta$, the real part of one of the eigenvalues (the one which associated with the rotational temperature $\theta_r$) can become negative for values of $\theta_t$ belonging to the range $\theta_t^{(1)}<\theta_t< \theta_t^{(2)}$. The critical values $\theta_t^{(i)}$ depend on $n^*$, $T_\text{ex}^*$, $\al$, $\beta$, and $\kappa$. This means that the steady simple shear flow solution is linearly \emph{unstable} in the region $\theta_t^{(1)}<\theta_t< \theta_t^{(2)}$.

As an illustration, Fig.\ \ref{fig8.1} shows the real part of the eigenvalues $\ell_i$ $(i=1,2,3,4)$ of the matrix $\mathsf{L}$ as a function of the translational temperature $\theta_t$ for $n^*=0.01$, $T_\text{ex}^*=1$, $\kappa=\frac{2}{5}$, $\al=1$, and $\beta=-0.5$. We find that two of the eigenvalues (let's denote them for instance by $\ell_2$ and $\ell_3$) are complex conjugate while the other two ($\ell_1$ and $\ell_4$) become a complex conjugate pair for high values of $\theta_t$. It is quite apparent that while the real part of $\ell_2$ (or $\ell_3$, since $\text{Re}\;\ell_2=\text{Re}\;\ell_3$) is always positive, the real parts of $\ell_1$ and $\ell_4$ become negative for certain critical values of $\theta_t$ (see the inset graph where the position of these critical values is more clearly shown). This means that there are two different unstable regions for this system.

The above feature is clearly confirmed in Fig.\ \ref{fig8.2} where we plot a phase diagram delineating the regions between stable and unstable solutions in the $\left\{\alpha,\theta_t\right\}$--plane for smooth inelastic hard spheres ($\beta=-1$) with
$n^*=0.01$, $T_\text{ex}^*=1$, and $\kappa=\frac{2}{5}$. While the hatched regions refer to values of $(\al,\theta_t)$ where the steady shear flow solution is \emph{stable}, the unfilled regions correspond to combined values of $\al$ and $\theta_t$ for which the steady solution is \emph{unstable}. It is worthwhile noticing that the dependence of the boundary line separating both stable and unstable regions on $\al$ is not quite trivial since at a given value of $\al$ there is a reentrance feature as the translational temperature $\theta_t$ increases: we first find a transition from the stable to unstable region, followed by a subsequent transition to the stable region. Surprisingly, the size of the unstable region decreases with inelasticity. As a complement of Fig.\ \ref{fig8.2}, Fig.\ \ref{fig8.3} shows two different phase diagrams in the $\left\{\beta,\theta_t\right\}$--plane for two values of the coefficient of normal restitution $\al$: $\al=0.9$ (panel (a)) and $\al=1$ (panel (b)). We observe first that there are two separate unstable regions around $\beta=-0.5$ in the case of $\al=1$. This is consistent with the findings of Fig.\ \ref{fig8.1}. The second unstable region corresponding to higher $\theta_t$'s is more squeezed than the first one. In addition, we see that the size of the unstable region decreases with increasing roughness ($\beta$ increases). This is more apparent in the case of the panel (a) of Fig.\ \ref{fig8.3} where only a single unstable region is found. This means that roughness attenuates the instability of the time-dependent sheared problem. In fact, at a given value of $\al$, there exists a critical value $\beta_c(\al)$ for which the unstable region is destroyed and hence, the steady solution is always linearly stable for $\beta>\beta_c$. In particular, $\beta_c \simeq 0.75$ for $\al=0.9$ and $\beta_c \simeq 0.94$ for $\al=1$. Figure \ref{fig8.3} also highlights the complex dependence of the boundary lines for $\al=1$ around $\beta=-0.5$ since the following series stable$\to$unstable$\to$stable$\to$unstable$\to$stable occurs when $\theta_t$ increases at fixed $\beta$.

In summary, our stability analysis shows that there are regions of the parameter space of the problem where the steady simple shear flow state can be linearly unstable. This restricts of course  the analysis performed here for rheology to specific regions of the parameter space where the steady solution is stable. Hopefully, the size of the stable regions is in general larger than that of unstable regions.

\section{Summary and discussion}
\label{sec9}

The determination of the non-Newtonian transport properties in inertial suspensions under simple shear flow has stimulated in the past few years the use of kinetic theory tools. Starting from the Boltzmann (which holds for very dilute systems) and/or the Enskog (which applies for moderately dense systems) kinetic equations, several works \cite{TK95,SMTK96,ChVG15,SA17,SA20,HTG17,HTG20,HT19,GGG19} have obtained explicit expressions of the shear-rate dependence of the kinetic temperature, the non-Newtonian viscosity, and the viscometric functions. In most of the cases, the analytical results have been validated against computer simulations showing in general good agreement for conditions of practical interest. An interesting conclusion is that the viscosity exhibits DST for very dilute systems; \cite{HT19,GGG19} this means that there is a sudden relative increase of viscosity with increasing shear rate. On the other hand, it has been also shown that DST becomes gradually into CST as the density increases.

However, all previous theoretical works\cite{TK95,SMTK96,ChVG15,SA17,SA20,HTG17,HTG20,HT19,GGG19} have considered inertial suspensions of \emph{smooth} inelastic hard spheres and hence, the effects of tangential friction in particle collisions on non-Newtonian rheology have been neglected. In the context of kinetic theory, we are not aware of any previous attempt on addressing the impact of roughness on the non-Newtonian transport properties. In this paper, we have addressed this problem; more specifically and due to the complexity of the problem we have considered a granular suspension of inelastic rough hard spheres at low density. In this case, the Boltzmann kinetic equation conveniently adapted for accounting the effect of the interstitial gas on grains is a reliable equation for obtaining the kinetic contributions to the temperature and the relevant elements of the pressure tensor.

In the case of smooth spheres, \cite{HTG17,HTG20,HT19,GGG19} the influence of the gas phase on solid particles has been usually accounted for by a gas-solid force constituted by two terms: (i) a drag force term proportional to the (instantaneous) velocity $\mathbf{v}$ plus (ii) a stochastic term represented by a Fokker--Planck operator of the form $-(\gamma_t T_\text{ex}/m)\partial^2 f/\partial v^2$. While the first term models the friction of grains on the continuous gas phase, the second one takes into account thermal fluctuations. On the other hand, when the spheres are not completely smooth and there is a certain friction between both spheres, one has also to take into account the coupling between the rotational degrees of freedom of grains and the gas phase. Here, we have assumed that this coupling has a similar structure to the one assumed in the smooth case and so, one has to add two new terms in the corresponding suspension model: a term proportional to the angular velocity $\boldsymbol{\omega}$ plus a Fokker--Planck operator of the form $-(\gamma_r T_\text{ex}/m)\partial^2 f/\partial w^2$. The coefficients $\gamma_t$ and $\gamma_r$ are proportional to the square root of the background temperature $T_\text{ex}$.

Once the suspension model is defined, as a first goal we have approximately solved it by Grad's moment method.\cite{G49} More specifically, we have evaluated the collisional moment $\mathcal{J}[V_k,V_\ell|f,f]$ [defined by Eq.\ \eqref{2.12}] by using the Grad's distribution \eqref{4.2}. The knowledge of this collisional moment allows us to obtain the explicit forms of the (reduced) rotational $\theta_r$ and translational $\theta_t$ temperatures as well as the (reduced) relevant elements of the pressure tensor $P_{k,\ell}^*$ in terms of the parameter space of the problem (the restitution coefficients $\al$ and $\beta$, the reduced moment of inertia $\kappa$, the reduced shear rate $a^*$, the reduced background temperature $T_\text{ex}^*$, and the reduced density $n^*$). Although the determination of non-Newtonian rheological properties (which are directly related with the second-degree velocity moments) is the most important objective of the present contribution, higher degree velocity moments are also relevant since they provide some indirect information on the velocity distribution function, specially in the high velocity region. Given that their evaluation from the true Boltzmann equation is quite intricate, as a second goal we have obtained them by considering a BGK-like kinetic model \cite{S11} recently proposed for inelastic rough hard spheres. Beyond non-Newtonian rheology, the fourth-degree moments are the first nontrivial moments in the steady simple shear flow problem. Their knowledge allows us to disclose partially the combined effect of the different physical mechanisms (shearing, gas phase, inelasticity) involved in the problem on the distribution function.

Regarding non-Newtonian rheology, the results derived here for inelastic rough hard spheres show no new surprises relative to the earlier works for smooth inelastic hard spheres:\cite{HTG17,HTG20,HT19,GGG19} the flow curve for the non-Newtonian viscosity $\eta^*(a^*)$ exhibits an $S$-shape and hence, DST is present. This means that $\eta^*$ discontinuously increases/decreases if $a^*$ is gradually increased/decreased [see the panel (c) of Fig. \ref{fig5}]. We have also observed that, at a given value of $\al$, the dramatic increase of viscosity is slightly mitigated by roughness (namely, as $\beta$ increases). The influence of roughness on rheology is more significant in the case of the (reduced) rotational temperature $\theta_r$. The panel (b) of Fig.\ \ref{fig5} highlights that, for large shear rates, $\theta_r$ increases with increasing $\beta$.

With respect to the fourth-degree moments, at a given value of the coefficient of normal restitution $\al$, surprisingly the BGK results show that the shear-rate dependence of those moments is practically independent of roughness in the range $a^*\leq 1$, where nonlinear effects are already important. This feature contrasts with the behavior of $\eta^*(a^*)$ since the value of $\eta^*$ clearly differs from its Navier--Stokes form in this range of values of the shear rate. For larger shear rates, we find that the fourth-degree moments also display an $S$-shape in a similar way to the viscosity $\eta^*$ (see Fig.\ \ref{fig7}).

As a complement of the previous results, we have also analyzed the stability of the steady simple shear flow solution for non-Newtonian rheology. To perform this analysis, since $\theta_t(a^*)$ is a multi-valued function of $a^*$, it is more convenient to take $\theta_t$ as an independent input parameter instead of the (reduced) shear rate. In this case, as happens for smooth spheres, \cite{HT19} the linear stability analysis shows regions of the parameter space of the system where the steady solution is linearly \emph{unstable}. More specifically, for given values of the set ($n^*$, $T_\text{ex}^*$, $\kappa$, $\al$), the steady solution becomes unstable in the region $\theta^{(1)}<\theta_t<\theta_t^{(2)}$, where the critical values $\theta_t^{(i)}$ depend on the coefficient of tangential restitution $\beta$. In addition, as the panel (b) of Fig.\ \ref{fig8.3} clearly illustrates, the dependence of the boundary lines delimitating stable/unstable regimes on $\beta$ is quite complex and in fact, there may be two or more separate unstable regions. It is worthwhile noticing that the unstable region usually belongs to the range of (reduced) shear rates where DST appears [see the thick lines of the panel (c) of Fig.\ \ref{fig5}]. Thus, it would be tentative to speculate on the possible relation between DST and instability, although this connection requires a more rigorous analysis. We plan to elucidate this point in the near future by considering a time-dependent inhomogeneous solution.

As mentioned in Sec.\ \ref{sec1}, the origin of DST has received a lot of attention in the past few years. Several mechanisms \cite{BJ14} have been proposed, most of them directly related to the complex structure of dense suspensions. On the other hand, as already discussed in Ref.\ \onlinecite{GGG19}, what is surprising here is the existence of DST in a structurally simple system. In this case, the origin of DST in dilute suspensions of inelastic hard spheres could be associated with both non-Newtonian rheology in far from equilibrium states as well as the effect of the interstitial gas on the dynamics of inelastic rough hard spheres.

\vicente{The fact that the roughness of spheres does not have a significant impact on DST (in the sense that the trends observed here are qualitatively similar to those observed for smooth spheres) could be in part due to the Fokker--Planck suspension model considered in this paper. As widely discussed in sec.\ \ref{sec2}, the above suspension model neglects the coupling between translational and rotational degrees of freedom of grains in the form of the operator $\mathcal{F}^\text{rot} f$. A way of accounting for this coupling in our theory would be to retain a term proportional to the vectorial product $\mathbf{v}\times \boldsymbol{\omega}$ in the form of $\mathcal{F}^\text{rot}$. This would necessarily give rise to new contributions in Grad's solution coming from the combination of traceless dyadic products of $\mathbf{V}$, $(\mathbf{V}\cdot \boldsymbol{\omega})$, and $\mathbf{V}\times \boldsymbol{\omega}$. The extension of the present theoretical results by considering the above terms in Grad's solution is a very challenging problem to be carried out in the future.}

It is apparent that the theoretical results presented here are relevant to make a comparison with computer simulations. Previous simulations \cite{HTG17,HTG20,HT19} carried for perfect smooth inelastic spheres ($\beta=-1$) have shown a good agreement with kinetic theory results, as is clearly illustrated in most of the plots presented along the paper. We expect that this agreement is also extended to the case of inelastic rough hard spheres. We plan to carry on those simulations in the near future. Another possible future project is the extension of the present results to finite densities by considering the Enskog kinetic equation. \vicente{In this context, an interesting question is to see if actually there is a transition from DST to CST as the density increases in a similar way as in the limit case of perfectly smooth spheres. Work on this line will be performed in the next future.}

\acknowledgments

This paper is dedicated to the memory of Prof.\ Jason Reese who made significant contributions to gas and liquid flows at the micro/nano scale. The present work has been supported by the Spanish Government through Grant No. FIS2016-76359-P and by the Junta de Extremadura (Spain) Grant Nos. IB16013 and GR18079, partially financed by ``Fondo Europeo de Desarrollo Regional'' funds. The research of Rub\'en G\'omez Gonz\'alez has been supported by the predoctoral fellowship BES-2017-079725 from the Spanish Government.

\appendix
\section{Navier--Stokes shear viscosity coefficient of dry granular gases}
\label{appB}

The explicit expression of the Navier--Stokes shear viscosity of a \emph{dry} gas of inelastic rough hard spheres is displayed in this Appendix. \cite{KSG14} It is given by
\beq
\label{b1}
\eta_\text{NS}=\frac{nT_t}{\nu_t}\frac{1}{\nu_\eta^*-\frac{1}{2}\zeta^*},
\eeq
where $\nu_t$ and $\nu_\eta^*$ are defined by Eqs.\ \eqref{4.9} and \eqref{4.16}, respectively, and the (reduced) cooling rate $\zeta^*$ is
\beq
\label{b2}
\zeta^*=\frac{5}{12}\frac{1}{1+\theta}\left[1-\al^2+(1-\beta^2)\frac{\kappa+\theta}{1+\kappa}\right].
\eeq
Here, the temperature ratio $\theta\equiv T_r/T_t$ is
\beq
\label{b3}
\theta=h+\sqrt{1+h^2},
\eeq
where $h$ defined by
\beq
\label{b4}
h=\frac{(1+\kappa)^2}{2\kappa(1+\beta)^2}\left[1-\al^2-(1-\beta^2)\frac{1+\kappa}{1+\kappa}\right].
\eeq


\textbf{DATA AVAILABILITY}

The data that support the findings of this study are available from the corresponding author upon reasonable request.



%

\end{document}